\documentclass[prx,amsmath,amssymb, notitlepage, twocolumn,
nofootinbib,
superscriptaddress,
longbibliography
]{revtex4-1}
\usepackage{caption}
\usepackage{subcaption}
\usepackage{amsmath}
\usepackage{tabularx,graphicx}
\usepackage{epstopdf}
\usepackage{graphicx}
 \usepackage{spectralsequences} 
\usepackage{latexsym}
\usepackage{amssymb}
\usepackage{tikz-cd}
\usepackage{amsmath}
\usepackage{color, colortbl}
\usepackage{psfrag}
\usepackage{bbm}
\usepackage{bm}
\usepackage{titlesec}
\usepackage{dsfont}
\usepackage{feynmp}
\usepackage{slashed}
\usepackage{multirow}
\usepackage{framed}
\newcommand{\beq}{\begin{eqnarray}}
	\newcommand{\eeq}{\end{eqnarray}}

\newcommand{\Z}{\mathbb{Z}}

\newcommand{\bsp}{\begin{split}}
	\newcommand{\esp}{\end{split}}

\newcommand{\sgn}{{\rm sgn}}

\usepackage{color}
\definecolor{darkblue}{rgb}{0.,0.,0.4}
\definecolor{darkred}{rgb}{0.5,0.,0.}
\definecolor{BlueViolet}{RGB}{138,43,226}
\definecolor{SkyBlue}{RGB}{30,144,255}
\definecolor{DarkGreen}{RGB}{0,100,0}
\usepackage[pdftex,colorlinks=true,linkcolor=darkblue,citecolor=blue,urlcolor=darkred]{hyperref}
\usepackage[normalem]{ulem}

\usepackage{amsthm}
\theoremstyle{plain}
\newtheorem*{theorem*}{Theorem}
\newtheorem{theorem}{Theorem}
\newtheorem{definition}{Definition}%[section]

\begin{document}
	\title{Average Symmetry-Protected Topological Phases}
	
	\author{Ruochen Ma}
	\affiliation{Perimeter Institute for Theoretical Physics, Waterloo, Ontario, Canada N2L 2Y5}
	\affiliation{Department of Physics and Astronomy, University of Waterloo, Waterloo, ON, N2L 3G1, Canada}

	\author{Chong Wang}
	\affiliation{Perimeter Institute for Theoretical Physics, Waterloo, Ontario, Canada N2L 2Y5}

\begin{abstract}

Symmetry-protected topological (SPT) phases are many-body quantum states that are topologically nontrivial as long as the relevant symmetries are unbroken. In this work we show that SPT phases are also well defined for \textit{average} symmetries, where quenched disorders locally break the symmetries, but restore the symmetries upon disorder averaging. An example would be crystalline SPT phases with imperfect lattices. Specifically, we define the notion of average SPT for disordered ensembles of quantum states. We then classify and characterize a large class of average SPT phases using a decorated domain wall approach, in which domain walls (and more general defects) of the average symmetries are decorated with lower dimensional topological states. We then show that if the decorated domain walls have dimension higher than $(0+1)d$, then the boundary states of such average SPT will almost certainly be long-range entangled, with probability approaching $1$ as the system size approaches infinity. This generalizes the notion of t'Hooft anomaly to average symmetries, which we dub ``average anomaly''. The average anomaly can also manifest as constraints on lattice systems similar to the Lieb-Schultz-Mattis (LSM) theorems, but with only average lattice symmetries. We also generalize our problem to ``quantum disorders'' that can admit short-range entanglement on their own, and develop a theory of such generalized average SPTs purely based on density matrices and quantum channels. Our results indicate that topological quantum phenomena associated with average symmetries can be at least as rich as those with ordinary exact symmetries.

\end{abstract}

\maketitle

\tableofcontents

\section{Introduction}
\label{sec:intro}

Symmetry classifies matter\cite{LandauLifshitz,McGreevy2022}. In the realm of symmetry-protected topological (SPT) quantum phases\cite{Pollmann2012,chenx2013,chen2013,Senthil2015}, symmetries are required in order to sharply distinguish different states of matter. For example, in topological insulators (TIs)\cite{HasanKane2010,QiZhang2011} the $U(1)$ charge conservation and time-reversal symmetries must be preserved to distinguish topological and trivial insulators. In more general terms, a nontrivial SPT state is short-range entangled (SRE) in the sense that it can be adiabatically connected to a trivial un-entangled product state, but such adiabatic paths are forbidden if the ``protecting symmetry'' is preserved. 

A natural question is, how exact does the protecting symmetry has to be? Specifically, if we have some \textit{quenched disorders} that locally break the protecting symmetry, but on average still respect the symmetry (such as magnetic impurities in TI), could the state still be in some topologically nontrivial phase? In other words, could \textit{average symmetry} protect nontrivial topological phases?

Previous studies have shown that some features of SPT phases in clean systems survive “statistically-symmetric" disorder present on the boundary, which lead to the concept of “statistical topological insulator" \cite{Fulga2012,Milsted2015}. One example is the three dimensional (3D) weak TI, made by stacking layers of a 2D TI. The surface state of the 3D weak TI is protected against Anderson localization even with strong disorder, if the translation symmetry along the stacking direction is restored by disorder averaging\cite{Ringel2012,Mong2012}. A similar delocalization appears on the surface of a 3D strong TI subject to a random magnetic field with zero mean\cite{Fu2012}, and even richer phenomena were discussed in the presence of interactions\cite{Altshuler2013,Chou2dTI,Chou2019,Chou2021,Kimchi2020}.

Key questions, however, remain unanswered. Previous studies focused on the effects of disorders on the boundary. It is then natural to ask: are the bulk topological phases sharply defined with average symmetries? If so, what are their signatures in the bulk, and how could we classify such phases? This question is particularly relevant for SPT phases protected by crystalline symmetries \cite{Fu2011,Song2017,Thorngren2016}, since impurities and lattice defects are ubiquitous in crystalline solids and the symmetry of an ideal lattice in reality is respected at best only on average. Of course when the material sample is of high quality, one can treat the ideal lattice as a good approximation. However, if the disorders become non-negligible, does the entire notion of crystalline SPT lose its meaning?

Even for the boundary physics, the problems were tackled on a case-by-case basis in existing literature. Given a symmetry group $\mathcal{G}$ that contains an average symmetry $G$, how do we systematically decide whether the boundary has to be nontrivial in some way? For free fermion states this issue was discussed in Ref.~\cite{Fulga2012}, but for interacting systems the question is largely open. In the standard theory of clean SPT, the non-triviality of the boundary is ultimately guaranteed by the quantum anomaly (more precisely, t'Hooft anomaly), which is decided by the bulk topological invariant. Familiar examples include the $(3+1)d$ TI surface protected by parity anomaly, and the integer quantum Hall edge protected by chiral anomaly. So the question about boundary properties can be rephrases as: does the notion of quantum anomaly exist for average symmetry? If anomaly can indeed be defined, how does such ``average anomaly'' constrain the infrared (IR) dynamics of the boundary theory?

Systems with t'Hooft quantum anomaly does not only appear on the boundary. An important class of such examples includes lattice systems with Lieb-Schultz-Mattis (LSM) constraints\cite{lieb1961two,Oshikawa2000,Hastings2003}. The most familiar example is a translation invariant lattice spin system with $SO(3)$ spin rotation symmetry and a spin-$1/2$ moment per unit cell. It is known that this system has a mixed t'Hooft anomaly between the discrete lattice translation and spin rotation symmetries\cite{Cheng2015}. As a consequence, the low energy dynamics cannot be trivial: either the symmetry will be spontaneously broken, or the system will form some long-range entangled ground state that is either topologically ordered or gapless. Similar t'Hooft anomalies also arise for other internal symmetries as long as they admit projective representations, and for other lattice symmetries such as rotation and reflection\cite{Po2017}. Recently, it was shown in Ref.~\cite{Kimchi2018} that, at least for $(1+1)d$ spin chains with $SO(3)$ symmetry, the LSM constraint holds even if the translation symmetry becomes only an average symmetry. Even though the argument in Ref.~\cite{Kimchi2018} does not make explicit connection to anomaly, it does suggest that the LSM anomaly should exist for arbitrary dimensions and for more general symmetries (such as time-reversal). Making this connection more explicit, precise and general is an open direction of great importance.

In this work we address all of the above issues. The key is to realize that 
\begin{enumerate}
    \item SPTs, whether in the bulk or on the boundary, are characterized by the properties of symmetry defects. The symmetry defects, e.g. twisted boundary conditions and gauge fluxes, may carry quantized invariants that can be used to define different phases. A well-known example is that in the bulk of a 3D TI, a unit magnetic monopole carries half-integral electric charge \cite{Qi2008,Rosenberg2010,Witten2016}.
    \item For an average symmetry $G$ in a disordered ensemble, even though the ground state in each disorder realization is not a $G$-eigenstate, we can still define defects associated with $G$ for the entire ensemble -- all we need is to modify the disorder potential accordingly. 
\end{enumerate}
Therefore by characterizing the defects, or equivalently domain walls of the average symmetries, we obtain an understanding of average SPT phases. Essentially, lower-dimensional states can be decorated onto the domain walls, similar to the case of clean SPTs. This ``decorated domain wall''\cite{chenlu2014} picture turns out to be powerful both in the bulk and on the boundary.

\subsection{ Summary of results}

We highlight the main results of this work below. This part will also serve as a map for the rest of this paper.

\begin{enumerate}
    \item In Sec.~\ref{sec:general}, we carefully define various basic notions. Specifically,
      \begin{itemize}
          \item We consider a disordered ensemble of local Hamiltonians $H_I$, where $I$ labels different disorder realizations. We consider quenched disorders that are spatially uncorrelated, up to some exponentially decaying tales.
          \item We consider two types of symmetries: an \textit{exact symmetry} is a symmetry for any disorder realization, namely $H_I$ is invariant under an exact symmetry for any $I$; an \textit{average symmetry}, in contrast, does not keep each individual $H_I$ invariant, but transforms $H_I$ to a different disorder configuration $H_{I'}$ with the same realization probability. We consider cases where the ensemble of bulk ground states $\{|\Psi_{I}\rangle\}$ does not spontaneously break any symmetry (exact or average).
          \item Similar to the study of SPT phases in clean systems, we demand each $H_I$ to have a gapped unique ground state $|\Psi_I\rangle$. Furthermore, we demand different realizations to be adiabatically connected to each other while preserving the exact symmetries: $|\Psi_{I}\rangle=U_T|\Psi_{I'}\rangle$, where $U_T$ represents a finite-time local Hamiltonian evolution, or equivalently a finite-depth local unitary circuit. Ground state ensembles that satisfy these conditions are dubbed \textit{short-range-entangled ensembles}. 
          \item We define two short-range-entangled ensembles to belong to the same \textit{average SPT} phase, if their Hamiltonians can be continuously tuned to each other while keeping the ground states symmetric and short-range-entangled.
      \end{itemize}
    
   % In particular, even though we do not assume the disorders to be weak, we do assume that the ground states in different disorder realizations to be adiabatically connected to each other. This, together with other physically intuitive assumptions, allow us to make controlled arguments throughout this work.
    
    \item In Sec.~\ref{sec:domain} we argue that a powerful way to think about average SPT phases is to study topological defects, such as domain walls and vortices, associated with the average symmetry. For an ensemble of preserves the average symmetry, these topological defects should proliferate at arbitrarily large length scale. To construct nontrivial average SPT states, we decorate the average symmetry defects with nontrivial topological phases protected by the exact symmetry. For example, if we introduce magnetic disorders in a three-dimensional strong topological insulator\cite{Fu2007}, we have two-dimensional domain walls of time-reversal breaking (introduced by the magnetic impurities) percolating throughout the bulk, and each domain wall is decorated with an integer quantum Hall states.  
    
    The decorated domain wall approach allows us to systematically classify average SPT phases. The simplest case is bosonic systems with total symmetry $\mathcal{G}=K\times G$, where $K$ is exact and $G$ is average: within the group-cohomology formalism, such average SPT phases are classified by (Theorem.~\ref{thm:DDWcoho})
    \begin{equation}
    \label{eq:decorIntro}
    \sum_{p=1}^{d+1}H^{d+1-p}(G,H^p(K,U(1))).
    \end{equation}
    In particular, group-cohomology states protected solely by $G$ (classified by $H^{d+1}(G,U(1))$) becomes trivial as $G$ becomes an average symmetry. This result also applies when $G$ is the average lattice symmetry, appropriate for realistic crystalline systems, and $K$ is the exact internal symmetry. The result can also be extended to states beyond group-cohomology. In Table~\ref{classtable} we list the classification of bosonic SPT phases for several simple symmetry classes, in space dimensions $1,2,3$, including those beyond group-cohomology. 
    
    \item In Sec.~\ref{sec:boundary} we show using a modified flux-insertion argument that, when nontrivial states are decorated on domain walls with dimensions higher than $(0+1)d$ (e.g. $p>1$ in Eq.~\eqref{eq:decorIntro}), the boundary state is almost certainly long-range entangled (or long-range correlated). More precisely, the probability for a sample (a single state in the ensemble) to have boundary correlation length $\xi$ exceeding any finite value $\xi_0$ approaches $1$: $P(\xi>\xi_0)\to 1$, as the system size $L\to\infty$ (Theorem~\ref{thm:boundaryanomaly}). We anticipate that these nontrivial boundary states will result in measurable signals such as low-temperature thermal conductance. In contrast, if states are decorated on $(0+1)d$ domain walls (e.g. $p=1$ in Eq.~\eqref{eq:decorIntro}), then the boundary state for all disorder realizations can be short-range entangled -- the only nontrivial feature in this case is that different disorder realizations may not be adiabatically connected in the presence of such mildly anomalous boundary. In Table~\ref{classtable} we list, in parenthesis, those states that do (almost certainly) have long-range entangled boundary states.
    
    \item The above result on average anomaly is used in Sec.~\ref{sec:LSM} to show a Lieb-Schultz-Mattis (LSM) constraint for systems with average translation symmetry, where each lattice unit cell contains a projective representation of the exact on-site symmetry (such as spin-$1/2$ moment for $SO(3)$ or Kramers' doublet for time-reversal). We argue that in such systems the ground state will almost certainly be long-range entangled (or long-range correlated), with probability approaching $1$ as the system size $L\to\infty$.
    
    \item In Sec.~\ref{sec:fermionicexample} we consider fermions systems. First in Sec.~\ref{sec:known} we discuss some free-fermion examples: weak and strong topological insulators, in two and three dimensions. As a demonstration of the power of our approach, we reproduce, in a simple manner, several nontrivial results from previous literature.  We then systematically consider $(3+1)d$ fermion systems in two symmetry classes: AII class ($U(1)\rtimes \mathcal{T}$ with Kramers' doublet fermions) and AIII class ($U(1)\times \mathcal{T}$). In both cases we consider average time-reversal symmetries. The AII class is relevant for electronic solids with spin-orbit interactions, with magnetic impurities that locally break time-reversal; the AIII class is relevant for quantum Hall plateaux transitions with average particle-hole symmetry. We show that for the AII case, the classification is reduced from $\mathbb{Z}_2^3$ in the clean case to $\mathbb{Z}_2^2$; for the AIII case, the classification is reduced from $\mathbb{Z}_8\times\mathbb{Z}_2$ to $\mathbb{Z}_4\times\mathbb{Z}_2$. All these nontrivial states have long-range entangled surface states with probability one, except for the $n=2$ state in the $\mathbb{Z}_4$ factor of AIII, in which the surface state can be short-range correlated for each individual disorder realization. The anomaly structure for the AIII case is consistent with numerical simulations on multi-component quantum Hall plateau transitions.
    
    \item In Sec.~\ref{sec:generalizeddisorder} we further generalize our problem to quantum disorders, where disorders are described by quantum mechanical degrees of freedom that can form nontrivial (but still invertible) many-body entanglement within themselves. This converts our problem to the study of SPT phases in mixed states -- a problem that has been recently studied\cite{deGroot2021arXiv} in the context of open quantum systems. We find that
    \begin{itemize}
        \item States protected solely by the average symmetry, including invertible states that do not need any symmetry, become trivial.
        \item Time-reversal symmetry always behave as an average symmetry.
        \item Bosonic SPTs described by elements in Eq.~\eqref{eq:decorIntro} are still nontrivial. For this statement, we give a careful justification in $(1+1)d$ in terms of the string order parameters, and give a plausibility argument in the more general cases.
    \end{itemize}
\end{enumerate}

 \begin{table*}[tttt]
%\begin{center}
\begin{tabular}{|>{\centering\arraybackslash}m{1.5in}|>{\centering\arraybackslash}m{1.2in}|>{\centering\arraybackslash}m{1in}|>{\centering\arraybackslash}m{1in}|}
\hline
{\bf Symmetry} &  {\bf $(1+1)d$} & {\bf $(2+1)d$} & {\bf $(3+1)d$} \\ \hline
$\mathbb{Z}_2^{(ave)}$ & $0$ & $0$ & $0$ \\ \hline
$(\mathbb{Z}_2^T)^{(ave)}$ & 0 & 0 & $\Z_2$ ($\Z_2$) \\ \hline
$\mathbb{Z}_2\times\mathbb{Z}_2^{(ave)}$ & $\Z_2$ ($0$) & $\Z_2^2$ ($\Z_2$) & $\Z_2^2$ ($\Z_2$) \\ \hline
$\mathbb{Z}_2\times (\mathbb{Z}_2^T)^{(ave)}$ & $\Z_2$ ($0$) & $\Z_2^2$ ($\Z_2$) & $\Z_2^3$ ($\Z_2^2$) \\ \hline
$\mathbb{Z}^T_2\times \mathbb{Z}_2^{(ave)}$ &  $\Z_2$ ($\Z_2$) & $\Z_2$ ($\Z_2$) & $\Z^3_2$ ($\Z_2^3$) \\ \hline
\end{tabular}
%\end{center}
\caption{Classification of bosonic average SPT phases in some symmetry classes, in space dimension $d=1,2,3$. The classification in parenthesis are those with long-range entangled boundary states.
}
\label{classtable}
\end{table*}%

We end with a discussion on open questions in Sec.~\ref{sec:discussion}.

\section{Generalities}
\label{sec:general}

Let us begin by introducing some useful concepts and physically defining our questions more precisely. To start, we consider a fixed lattice Hilbert space with a local tensor product structure $\mathcal{H}=\otimes_i\mathcal{H}_i$ ($i$ labeling lattice sites), and an ensemble of \textit{local} Hamiltonians $\{H_I\}$ and their ground states $\{|\Psi_I\rangle\}$, with probability $\{P_I\}$. For concreteness the Hamiltonian takes the form
\beq
\label{eq:Ham}
H_I=H_0+\sum_i(v^{I}_i\mathcal{O}_i+h.c.),
\eeq
where $v^{I}_i$ is a quenched disorder potential ($I$ labeling a particular realization and $i$ labeling a lattice site), $\mathcal{O}$ is a local operator, and $H_0$ is the disorder-free part of the Hamiltonian. We require the disorder to be at most short-range correlated, namely $\overline{v^*_iv_j}$ should decay exponentially with $|i-j|$.

We now consider two types of global symmetries. The \textbf{exact symmetry} $K$ commutes with both the disorder-free part and the disordered part of the Hamiltonian, for any individual realization of the disorder. The \textbf{average symmetry} (or statistical symmetry)\cite{Fulga2012,Fu2012,Kimchi2018,Kimchi2020} $G$ only commutes with $H_0$ and is broken by each realization of the disorder potential, so effectively the disorder potential $v$ transforms non-trivially under $G$ {($g\in G: v\to g\cdot v$)}. We then require that the probability distribution $P[v_I]\equiv P_I$ to be invariant under a $G$ transform ($P[g\cdot v]=P[v]$), so the entire statistical ensemble stays invariant, hence the name average (or statistical) symmetry. For example, a random magnetic field $h(x)$ is odd under time-reversal $\mathcal{T}:h\to -h$, and time-reversal would be an average symmetry if $P(h)=P(-h)$. In real materials, the most common examples of average symmetry are crystalline symmetries such as lattice translation: in each sample the symmetries are broken by impurities and lattice defects. It is typically the case that the impurities and defects will appear randomly at different positions with equal probability, which makes the lattice symmetries valid on average. However, to keep the disorder potentials short-range correlated, we do need to keep the impurities and defects dilute so they do not interact with each other (the strength of impurity scattering potential is unrestricted on the other hand). For internal symmetries such as time-reversal, a natural way to effectively obtain average-symmetric disorder is to start with an exact symmetry, then break it spontaneously but with the order parameter varying randomly in space -- the most famous example of this kind is a spin glass order (for a concrete example see Ref.~\cite{Chou2dTI}).

For simplicity we will often focus on cases where the full symmetry of the ensemble $\mathcal{G}$ is given by $K\times G$. But we note that in general, $\mathcal{G}$ is given by the group extension, 
\begin{equation}
\begin{tikzcd}
1 \arrow[r] & K \arrow[r] & \mathcal{G} \arrow[r] & G \arrow[r] & 1,
\end{tikzcd}    
\label{eq:fullsymmetry}
\end{equation}
where $K\subset \mathcal{G}$ is a normal subgroup. $\mathcal{G}$ may or may not contain an anti-unitary (time reversal) element. We shall also assume that both $K$ and $G$ acts locally (namely their actions within each lattice unit cell are unentangled), so they do not suffer from any t'Hooft anomaly -- we shall come back to this issue later when discussing boundary properties.

We now proceed to define the analogue of symmetric short-range entangled (SRE) states, but for the entire statistical ensemble $\{H_I,|\Psi_I\rangle,P_I\}$. It is natural to consider cases in which each individual $|\Psi_I\rangle$ is SRE (and symmetric under $K$), namely each $H_I$ is gapped with a unique symmetric ground state. However for our purpose this is not enough: we would like to forbid the ensemble from containing states in different SRE phases (possibly protected by $K$) separated by topological phase transitions\footnote{If the disorder potential $v$ takes continuous values in a connected space, this would be automatically forbidden by imposing symmetric SRE on each individual state. However in more general cases, the condition has to be imposed separately.}. We therefore have

\begin{definition}
\label{def:SREensemble}
A $K$-\textbf{symmetric SRE ensemble} is one that only contains $K$-symmetric SRE ground states $\{H_I,|\Psi_I\rangle\}$, with any pair of states being adiabatically connected to each other while preserving $K$.
\end{definition}

Notice that we impose the symmetric SRE condition on \textit{all} states in the ensemble, including those rare states with vanishing probability in thermodynamic limit. This is to avoid potential subtleties from Griffiths-like singularities. We expect this no-rare-region restriction to be physically reasonable far away from quantum phase transitions. The interplay between rare region effects and topological phase transitions is a fascinating subject that we leave to future studies.

To study symmetric SRE states, we further demand that the ensemble of states $\{|\Psi_I\rangle\}$ does not break the symmetries spontaneously. For exact symmetries this simply means that each individual state $|\Psi_I\rangle$ does not break the symmetries (i.e. is not a cat state), which is anyway guaranteed by the symmetric SRE condition. The question is slightly subtler for average symmetries. One could, for example, detect spontaneous breaking of an average symmetry $G$ by measuring the average magnitude of the integrated order parameter
\beq
M\equiv\overline{|\sum_i\langle\phi_i\rangle|},
\eeq
where $\phi_i$ is some local order parameter (defined near site $i$) that transforms nontrivially under $G$, $\langle...\rangle$ denotes the expectation value with respect to a particular quantum state, and the overline denotes the disorder (ensemble) averaging (we shall use this notation throughout this paper). If the ensemble spontaneously breaks $G$, we expect $M$ to be proportional to the volume $L^d$. For symmetric ensembles we expect a smaller scaling -- for example, a trivial paramagnetic state will have $M\sim O(L^{d/2})$.

The above way of detecting spontaneous breaking of average symmetries, however, is not very convenient for our purpose. In this work, instead, we will guarantee the absence of spontaneous $G$-breaking primarily through proliferation of domain walls (or other defects such as vortices). Essentially, we demand that at sufficiently large length scale (larger than the correlation length), domain walls (or other appropriate defects) will always appear to restore the statistical $G$ symmetry. At low enough dimensions ($2d$ for discrete symmetries), such domain wall proliferation is always guaranteed by the Imry-Ma theorem\cite{ImryMa1975}.

Next we shall define the notion of \textbf{continuous symmetric deformation} -- the analogue of symmetric adiabatic evolution -- for our SRE ensembles. This task is relatively straightforward: we continuously deform both the Hamiltonians ($H_0$ and $\mathcal{O}_i$ in Eq.~\eqref{eq:Ham}) and the probability distribution of the disorder $P[v]$, such that (1) both the Hamiltonians and the disorder correlations remain short-ranged, and (2) the ensemble of states remains symmetric and SRE throughout.

We are now ready to define the notion of average SPT phases: 
\begin{definition}
\label{def:ASPT}
Two SRE ensembles, with exact symmetry $K$ and average symmetry $G$, belong to the same \textbf{average SPT phase} iff there is a path of continuous symmetric deformation connecting the two.
\end{definition}

As in most other topology problems, it is impractical to check all continuous paths between two states. Instead it is much more useful to construct topological invariants to distinguish different phases. This will be the task of next Section.

\section{Decorated domain wall approach}

\label{sec:domain}

In this section we will generalize the decorated domain wall approach\cite{chenlu2014}, a powerful construction for standard SPT phases, to the study of average SPT phases. Let us first review the idea of constructing standard SPT phases by the decorated domain wall (more generally, symmetry defect) construction in clean systems\cite{chenlu2014}, where all symmetries are exact. Starting from a phase in $(d+1)$-dimensional space-time, in which $G$ is broken spontaneously, a symmetric state can be obtained from condensation of $G$-domain walls. Non-trivial SPT phases are produced by decorating codimension $p$ (with respect to the space-time) topological defects of $G$ with $(d-p+1)$-dimensional SPT phases protected by the unbroken symmetry $K$ before the domain wall proliferation. In order for the condensation of $G$-domain walls to be gapped with a unique ground-state in the bulk, there is a set of consistency conditions for the defect decoration \cite{wang2021}, such that $G$-defect of each codimension is free of
$K$-anomaly. In this scheme the protected surface states appear naturally: topological defects that end at the surface carry the non-trivial boundary modes of the lower dimensional SPT phases protected by the symmetry $K$.

For simplicity let us tentatively assume $G$ to be discrete and unitary -- the more general cases are similar in conclusions but more subtle in details. The decorated domain wall approach can be equivalently formulated as follows: consider the SPT state $|\Psi\rangle$, and act on it with the symmetry element $g\in G$, but only in a (large enough) subregion $A$ (say with a disk geometry): $U_g^{A}\equiv\prod_{i\in A}U_g^{i}$ ($U_g^i$ is the local $g$-generator). The symmetric SRE nature of $|\Psi\rangle$ implies that acting with $U_g^{A}$ has nontrivial effect only near the boundary of $A$ \cite{Pollmann2010,Turner2011,Fidkowski2011}, namely $U_g^A|\Psi\rangle=V_g^{\partial A}|\Psi\rangle$, where $V_g^{\partial A}$ is a unitary operator that is nontrivial (non-identity) only near the boundary $\partial A$. In this case a decorated domain wall simply means that $V_g^{\partial A}$ creates a nontrivial phase in one dimension lower. Similar considerations can be carried out for defects with higher codimensions\cite{wang2021} and for anti-unitary symmetries\cite{Shapourian2017}.

Now we add quenched disorder that breaks the $G$-symmetry to the system. One can imagine that now the system consists of patches with different symmetry breaking patterns, which are pinned by the symmetry-violating disorder. In two adjacent patches, the states are related by an action of the broken symmetry. As a result, the interface between two adjacent patches naturally realizes a $G$-domain wall. The idea is that, similar to the case in clean systems, we can decorate the domain walls with nontrivial lower dimensional invertible phases, such as SPT phases protected by the exact symmetry. When the disorder has a random distribution so that the $G$-symmetry is restored on average, one again gets a $G$-defect network, extending over the entire system. 

The above picture is very similar to the standard (exact symmetry) SPT, but with one important difference: for standard SPT the domain walls proliferate as coherent quantum superpositions, with well defined phase factor associated with each domain wall configuration -- for example, the bosonic $\mathbb{Z}_2$ Levin-Gu state\cite{LevinGu2012} has a $(-1)$ factor for each Ising domain wall; for average SPT, however, the domain walls proliferate through classical probability, with no analogue of quantum phase factors. Therefore SPT states that are nontrivial due to such phase factors in the domain wall superpositions have no analogue in average SPT.

Let us try to make the decorated domain wall picture for average SPT more precise. Consider a particular realization, say, $\{v_i\}$ in Eq.~\eqref{eq:Ham}, with the ground state $|\Psi\rangle$. Now consider a different realization, with $\tilde{v}_i=v_i$ for $i$ outside of a region $A$, and $\tilde{v}_i=gv_ig^{-1}$ inside the region $A$ (for example for Ising symmetry $\tilde{v}_i=-v_i$ for $i\in A$), and denote the corresponding ground state as $|\tilde{\Psi}\rangle$. These two disorder realizations have essentially identical probability. The absence of spontaneous $G$-breaking, together with the SRE nature of the ensemble, implies that for large enough $A$, the two should look identical deep inside $\bar{A}$, and should differ only by a $g$-action deep inside $A$ -- the only potential nontrivial difference can only happen near the boundary $\partial A$. Formally,
\beq
\label{eq:DDW}
U^A_g|\Psi\rangle=V_g^{\partial A}|\tilde{\Psi}\rangle,
\eeq
with $V_g^{\partial A}$ defined nontrivially only on $\partial A$. If we choose a different disorder realization, say $v'_i$ with ground state $|\Psi'\rangle$, we can similarly define $\tilde{v}'_i$ and $|\tilde{\Psi}'\rangle$. By assumption (Def.~\ref{def:SREensemble}), all these states are connected through some $K$-symmetric adiabatic evolutions (or $K$-symmetric finite depth unitary circuit). Moreover, the evolution connecting $|\Psi\rangle$ to $|\Psi'\rangle$ (call it $W$) and that connecting $|\tilde{\Psi}\rangle$ to $|\tilde{\Psi}'\rangle$ (call it $W'$) must be identical deep inside $\bar{A}$ and differ only by conjugating $g$ deep inside $A$. These facts are enough to show that 
\beq
U^A_g|\Psi'\rangle=(V_g^{\partial A})'|\tilde{\Psi}'\rangle,
\eeq
where $(V_g^{\partial A})'$ and $V_g^{\partial A}$ only differ by an adiabatic evolution on $\partial A$. In other words, the topological nature of $V_g^{\partial A}$ does not depend on the choice of disorder realization, even though non-universal properties of $V_g^{\partial A}$ certainly does depend on details of the disorder potential. Similarly, one can show that the topological nature of $V_g^{\partial A}$ also does not depend on the choice of the region $A$, as long as $A$ is large enough -- essentially, Def.~\ref{def:SREensemble} requires different domain wall configurations to be adiabatically connected to each other, which in turn requires the decorating phases on the domain walls to remain the same no matter where the domain walls move to. The only way to change the topological property of $V_g^{\partial A}$ is to go through a phase transition -- at least for some of the states in the ensemble.

The above arguments establishes the (topological part of) $V_g^{\partial A}$, the ``decoration'' on the domain walls, as a robust property describing the corresponding average SPT phases.

\subsection{Topological response from replica field theory}
\label{sec:replicatheory}

Similar to the standard SPT theory, the decorated domain wall construction can be rephrased as a topological response theory for background gauge fields. For this purpose, we work with the path integral and use the replica trick: we replicate the Lagrangian $N_r$ times to obtain the action
\beq
\label{eq:replicaS}
S=\int dtd^dx\sum_{\alpha=1}^{N_r}\mathcal{L}[\phi_{\alpha}(x,t),v(x)]+\int d^dx V[v(x)],
\eeq
where $\phi$ represent all the dynamical degrees of freedom, $\alpha$ is the replica index, $v$ is the disorder potential. The first term represents the dynamics of $\phi$'s and their interactions with $v$, and the second term generates the classical probability of the disorder potential. Note that while the dynamical fields $\phi(x,t)$ depend on both space and time, the disorder potential $v(x)$ only varies in space and is constant in time. 

The replicated action Eq.~\eqref{eq:replicaS} is, by definition, invariant under the full symmetry group $\mathcal{G}$. The disorder potential $v$ is invariant under the exact symmetry $K$, but transforms non-trivially under the average symmetry $G$. There is no obstruction in coupling the theory in Eq.~\eqref{eq:replicaS} to a background gauge field in $\mathcal{G}$, call it $A^{\mathcal{G}}$. The only subtlety is that since $v$ is constant in time, any gauge transformation associated with $G$ must be constant in time. This then requires the $G$ gauge field, denoted as $A^G$, to be trivial along the time direction. Since the time component of a gauge field couples to the symmetry charge, the constraint on $G$ gauge field is simply a reflection of the fact that $G$-charge is not conserved for our system, and one cannot use $G$-charges to distinguish different phases of matter. The spacial components of $A_G$, on the other hand, are not constrained. In fact, the average symmetry defects, discussed in the decorated domain wall construction, can be precisely described using the spacial components of $A_G$ following standard procedures. For example, a nontrivial holonomy of $A^G$ along a spacial cycle represents a twisted boundary condition for both the dynamical fields and the disorder potential.  

We can now formally integrate out the dynamical fields $\phi$, and obtain the partition function that depends on the background gauge field $A^{\mathcal{G}}$ and the spacetime $(d+1)$-manifold $X$. For an invertible phase (such as SPT) in a clean setup, the global properties are included in a topological quantum field theory (TQFT) as the imaginary phase of the Euclidean partition function\cite{Kapustin2014},
\begin{equation}
    \mathrm{ln}(Z[X,\, A])\sim i S_{\mathrm{top}}[X,\, A]+\cdot\cdot\cdot,
    \label{eq:effectiveaction}
\end{equation}
in which the terms omitted are irrelevant below the bulk energy gap. In the presence of quenched randomness, the disorder-averaged effective action can be obtained from the replica limit \cite{altland2010condensed}
\begin{equation}
\begin{split}
   S[X,A^{\mathcal{G}}] &= \overline{\mathrm{ln}Z[X,A^{\mathcal{G}}]} \\
   &=\lim_{N_r\to 0} \frac{1}{N_r}\overline{ (Z[X,A ^{\mathcal{G}}]^{N_r}-1) },
      \end{split}
\label{eq:averageeffectiveaction}
\end{equation}
where the overbar denotes the disorder average. Analogous to clean systems, the topological term that survives the replica limit $N_r\to 0$ in Eq.~(\ref{eq:averageeffectiveaction}) encodes the topological properties of the disorder system.

Here we make a side remark. The maximal symmetry group (let us denote it by $\Tilde{\mathcal{G}}$) of the actions in Eq.~(\ref{eq:replicaS}) is not the “full symmetry group" $\mathcal{G}$ that acts diagonally on all replicas. For example, if $\mathcal{G}=K\times G$, then $\tilde{\mathcal{G}}=K^{N_r}\times G$, since each replica can transform under $K$ independently while leaving the Lagrangian invariant. More generally, $\Tilde{\mathcal{G}}$ is defined by the following morphism of short exact sequences:
\begin{equation}
\begin{tikzcd}
1 \arrow[r] & K^{N_r} \arrow[r]                    & \mathcal{G}^{N_r} \arrow[r]                  & G^{N_r} \arrow[r]          & 1 \\
1 \arrow[r] & K^{N_r} \arrow[r] \arrow[u, "\cong"] & \tilde{\mathcal{G}} \arrow[r] \arrow[u, "f"] & G \arrow[r] \arrow[u, "F"] & 1
\end{tikzcd}
\end{equation}
Here $F$ is the diagonal map, and we ignore the (in general discrete) rotation symmetry among the replicas. Coupling the overall $\Tilde{\mathcal{G}}$ symmetry to backgrounds enables us to calculate quantities such as 
\begin{equation}
    \lim_{N_r\to 0} \overline{\langle O_1^\alpha O_2^\beta \rangle}-\overline{\langle O_1^\alpha \rangle \langle O_2^\beta \rangle} ,
\end{equation}
where $\alpha \ne \beta$ and $O_1$, $O_2$ are arbitrary operators. Physically these quantities encode nontrivial sample-to-sample fluctuations in the disordered ensemble. Since we have assumed (Def.~\ref{def:SREensemble}) that different disorder realizations are adiabatically connected, we do not expect such sample-to-sample fluctuations to play an important role in our discussions. One can, however, ask whether by relaxing our assumptions, we can discover nontrivial topological properties associated with sample-to-sample fluctuations (such as a topological analogue of the universal conductance fluctuation\cite{LeeStone1985}), as may be captured by coupling to $\tilde{\mathcal{G}}$ gauge field. We leave this intriguing possibility for future study.

We are now ready to classify and characterize a large class of ASPT phases. For SRE phases for which the ground-state is unique and gapped on any closed spatial manifold, this problem is equivalent to the classification of associated invertible TQFTs (see Eq.~(\ref{eq:effectiveaction}) and Eq.~(\ref{eq:averageeffectiveaction})), studied by the cobordism theory \cite{Kapustin2015,Kapustin2014,Kapustin2014boson,Freed2016}. The static disorder modifies the classification by constraining the space-time configurations of the background fields of average symmetries, i.e. it quenches the holonomy of $A^G$ along the time cycle. For example, a clean topological phase remains nontrivial if and only if the corresponding TQFT remains non-trivial given this constraint. Next, with some simple examples, we illustrate how the topological response theory naturally leads to the decorated domain wall construction.

\subsection{Simple examples:\\ group cohomology states with $\mathcal{G}=K\times G$}

Let us consider bosonic SPT phase described by group cohomology\cite{chen2013}. For simplicity we also assume $\mathcal{G}=K\times G$, namely the group extension Eq.~(\ref{eq:fullsymmetry}) is trivial. In this case the group cohomology classification in $(d+1)$-dimensional space-time can be rewritten by the Künneth formula,
\begin{equation}
    H^{d+1}(\mathcal{G},U(1))=\sum_{p=0}^{d+1}H^{d+1-p}(G,H^{p}(K,U(1))),
    \label{eq:Künneth}
\end{equation}
in which the corresponding coefficient group is twisted if $G$ or $K$ contains time reversal \cite{Kapustin2014boson}. This mathematical formula can be understood from the perspective of topological effective actions. For a bosonic system, the topological action in Eq.~(\ref{eq:effectiveaction}) can be expressed as an integral of a local Lagrangian $\mathcal{L}$ over space-time,
\begin{equation}
    S_{\mathrm{top}}=2\pi\int_X \mathcal{L},
\end{equation}
where $\mathcal{L}$ is a $(d+1)$-dimensional cocycle, built out of flat background gauge fields (and $w_1(TX)$, the first Stiefel-Whitney class of the space-time, which can be viewed as the time reversal gauge field). In particular, $\mathcal{L}$ may be written as a cup-product $\mathcal{L}=\mathcal{L}_1\cup \mathcal{L}_2$, where $\mathcal{L}_1$ and $\mathcal{L}_2$ are two cocycles constructed from the background gauge fields in the theory, whose degrees sum to $(d+1)$. As the effective action of an SRE phase, we require $\mathcal{L}$ to be gauge invariant on any closed space-time. For a compact $X$, the Poincare duality $H^p(X)\cong H_{d+1-p}(X)$ (with the coefficient in any ring) enables us to rewrite the action as
\begin{equation}
    S_{\mathrm{top}}=2\pi \int_{\hat{\mathcal{L}}_2} \mathcal{L}_1,
    \label{eq:domainaction}
\end{equation}
where $\hat{\mathcal{L}}_2$ is the Poincare dual (with respect to $X$) of the cocycle $\mathcal{L}_2$. 

To make connections between current discussion and the decorated domain wall picture, note that if the cocycle $\mathcal{L}_2$ is taken to be the background gauge field $A^G$ of a symmetry $G$, the Poincare dual surface $\hat{\mathcal{L}}_2$ is simply a $G$-domain wall. The action in Eq.~(\ref{eq:domainaction}) hence describes an effective $(d-1+1)$-spacetime-dimensional topological phase, living on the wall. This precisely corresponds to the SPT phase decorated on the codimension-1 $G$-domain wall. Gauge invariance of $\mathcal{L}$ ensures the consistency between the decoration and the fusion rules of the domain walls. The same argument also holds for defects of higher codimensions, where $\mathcal{L}_2$ are cocycles of higher degrees built out of $A^G$. Therefore a physical interpretation of an element of $H^{d+1-p}(G,H^{p}(K,U(1)))$ in Eq.~(\ref{eq:Künneth}) is a consistent decoration of a $p$-dimensional $G$-defect by a $K$-SPT phase in $p$-dimensional space-time.

When static randomness turns $G$ into an average symmetry, $A^G$ can have non-trivial holonomies only around spatial cycles. Equivalently, this means the symmetry defect $\hat{\mathcal{L}}_2$ extends along time, while its spatial position is pinned. A straightforward observation is that the topological action of the $p=0$ element (i.e. the group cohomology $H^{d+1}[G,U(1)]$, with twisted coefficient if $G$ contains time reversal) in Eq.~(\ref{eq:Künneth}) becomes trivial if the holonomy of $A^G$ around time cycle is quenched, i.e. $\int_\tau A^G=0$. For example, the Levin-Gu state\cite{LevinGu2012} has partition function $S=\pi\int a\cup a\cup a$ ($a\in H^1(X,\mathbb{Z}_2)$ being the background $\mathbb{Z}_2$ gauge field), and is trivial if $a$ along the time direction is set to zero. This confirms our physical expectation based on the domain wall proliferation picture at the begining of Sec.~\ref{sec:domain}.

In contrast, the effective actions of elements in Eq.~(\ref{eq:Künneth}) with $p>0$ remain non-trivial. The physical picture is precisely the decorated domain walls discussed at the beginning of Sec.~\eqref{sec:domain}. We therefore conclude
\begin{theorem}
\label{thm:DDWcoho}
Bosonic SPT phases with symmetry $\mathcal{G}=K\times G$ ($K$ being exact and $G$ being average) described within group cohomology are classified by
\begin{equation}
\label{eq:DDWcoho}
    \sum_{p=1}^{d+1}H^{d+1-p}(G,H^p(K,U(1))).
\end{equation}
\end{theorem}
Namely, only mixed topological response between $G$ and $K$ (or pure response for $K$ alone) remain nontrivial as $G$ becomes average symmetry.

Going beyond group-cohomology classification, at least for bosonic systems with $\mathcal{G}=K\times G$, is not too complicated. The only new ingredient is that on the domain walls we can also decorate nontrivial invertible topological phases, resulting in mixed ``gauge-gravity'' topological response\cite{WangGuWen2015} -- essentially the cocycles in Eq.~\eqref{eq:domainaction} can also involve characteristic classes of the spacetime itself. Such response will remain nontrivial as $G$ becomes average symmetry. For example, decorating the chiral $E_8$ state\cite{Kitaev2006} on the average time-reversal domain walls results in the so-called $efmf$ state\cite{Vishwanath2013,Wang2013,Burnell2013} in $(3+1)d$ protected by the average time-reversal symmetry. So we conclude that as $G$ becomes average symmetry, a nontrivial $K\times G$ boson SPT in the clean limit becomes trivial iff the SPT is characterized by a pure $G$-gauge response (no $K$ gauge field or gravity involved).

In Table~\ref{classtable} we list the classification of bosonic SPT phases for some simple symmetry classes, in space dimensions $1,2,3$, including states beyond group-cohomology. 

So far we have analyzed clean SPT phases and showed many of those remain nontrivial as $G$ becomes average symmetry. From a different perspective, however, in our analysis we have exhausted all the possible ways to decorate the domains walls. Therefore what we obtained is also a complete classification of average SPT phases, at least within the decorated domain wall picture. We note that, once we go beyond the simple case of $\mathcal{G}=K\times G$, the completeness of this classification is no longer guaranteed, and we shall explore this extremely intriguing possibility in future study.

\subsection{Application: crystalline SPT}

For the purpose of classifying SPT phases, crystalline symmetries can be treated with internal symmetries on the same footing. This fact, known as the ``crystalline equivalence principle''\cite{Thorngren2016}, allows us to apply results in this section to crystalline SPT phases in realistic crystals, where the lattice symmetries are only preserved on average.

Many physically relevant examples can be described as $\mathcal{G}=K\times G$, where $K$ is an exact internal symmetry (e.g. time-reversal, spin $SO(3)$ etc.), and $G$ represents the lattice symmetries such as translation, rotation and reflection. For boson (or spin) systems, Eq.~\eqref{eq:DDWcoho} then gives the group-cohomology classifications. States beyond group-cohomology are classified similarly, with nontrivial invertible states decorated on domain walls.

We can also give a simple example of crystalline SPT phase that is nontrivial in the clean limit, but becomes trivial once the crystal symmetry becomes average symmetry: consider a $(2+1)d$ SPT state with $C_2$-rotation (inversion) symmetry. The only nontrivial state, which is the crystalline counterpart of the Levin-Gu state\cite{LevinGu2012}, can be constructed by putting a nontrivial $C_2$ charge at the $C_2$ rotation center. When the $C_2$ symmetry becomes average, the notion of ``nontrivial $C_2$ charge'' no longer makes sense, therefore the state becomes trivial.

It is illuminating to compare the decorated defect pictures in three differnt types of SPTs:
\begin{itemize}
    \item In the standard internal symmetry SPTs, the defects condense into quantum superpositions. In other words, the defects proliferate quantum mechanically.
    \item In crystalline SPTs, the crystalline defects, such as crystalline unit cells, rotation axes and reflection planes, are static\cite{Song2017,Huang2017,Song2018b,ElseThorngren2019,Zhang2022}.
    \item In average SPTs, the average symmetry defects are static in each disorder realization, but they proliferate probabilistically in the ensemble of states.
\end{itemize}
In this sense, the average SPTs are somewhat in between internal and crystalline SPTs. As we have discussed in this Section, although subtle distinctions between average and standard SPTs do exist, the overall pictures are quite similar in terms of decorated and proliferated defects (domain walls).

\subsubsection{ Example: mirror-symmetric topological crystalline insulator}
\label{sec:mTCI}

As an example, we present a detailed discussion of an average mirror-symmetric topological crystalline insulator\cite{Teo2008}. Generalizations to other average crystalline SPT phases will be straightforward.

We consider a three-dimensional free fermion insulator with charge $U(1)$ and mirror reflection $\mathcal{M}_z:z\to-z$. The simplest nontrivial state has one Dirac cone on the surface. Within band theory, the bulk state is characterized by a ``mirror Chern number''\cite{Teo2008} $n_{\mathcal{M}}=1$. Furthermore, it has been demonstrated\cite{Isobe2015} that this state remains nontrivial even with interactions as long as the gap remains open during the adiabatic turning-on of the interactions. In real space, this state can be adiabatically deformed to a particularly simple limit\cite{Song2017}: on the mirror reflection plane $z=0$, we decorate an integer quantum Hall (IQHE) state with $\sigma_{xy}=1$, and on the other reflection-invariant plane at $z=L/2\sim -L/2$ (where $L$ is the system size, and periodic boundary conditions are assumed), we decorate an opposite IQHE state with $\sigma_{xy}=-1$. In the absence of mirror symmetry, the state can be adiabatically trivialized through a generalization of the Thouless pump\cite{Thouless1983}, which pumps the IQHE state at $z=0$ to $z=L/2$ through one side of the bulk (say the right side). Specifically, the pump is an adiabatic deformation on the right side of the system $H_{z>0}(\tau)$ from time $\tau=0$ to $\tau=T$ ($T$ does not scale with $L$), such that away from $z=0$ and $z=L/2$ we have $H(T)=H(0)$, and at $z=0$ and $z=L/2$ the IQHE layers are eliminated at $\tau=T$. The left side of the system cannot undergo the same pump because it would add another pair of IQHE layers to the reflection-invariant planes. As such, adiabatic pumping is not allowed with exact mirror symmetry.

When mirror symmetry is preserved only on average, it requires the probability distribution of the Hamiltonians to be mirror-symmetric: $P[H_{z>0}]=P[H_{z<0}]$. We now argue that, even with average mirror symmetry, the above adiabatic pumping is still not possible without violating the basic assumptions in Sec.~\ref{sec:general}.  For the disordered ensemble, the adiabatic deformation from $\tau=0$ to $\tau=T$ is specified for each disorder realization $I$: $H^I(\tau)=\sum_i H^I_i(\tau)$, where $H^I_i$ is the local Hamiltonian density around site $i$ for realization $I$, and by assumption the probability distributions of $H_i$ at different $i$ are uncorrelated up to exponential tales (Sec.~\ref{sec:general}). Suppose we have one realization $I_0$ in which $H^{I_0}_{z>0}(\tau)$ pumps away the IQHE layers at $z=0$ and $z=L/2$, and $H^{I_0}_{z<0}(\tau)$ does not pump anything nontrivial. The average mirror symmetry, together with the short-range-correlated nature of the disorders, tell us that for any $z_0>0$ there must be another disorder realization $I_1$, in which $H^{I_1}_{z<0}$ and $H^{I_1}_{z_0<z<L/2}$ do not pump anything, while $H^{I_1}_{0<z<z_0}$ pumps a pair of IQHE to $z=0$ and $z=z_0$. This means that, at the final time $\tau=T$, $H^{I_1}_{z\approx z_0}$ has a nontrivial IQHE ground state, while $H^{I_0}_{z\approx z_0}$ has a trivial ground state (similar conclusion also holds for any $z_0<0$). Therefore the total Hall conductivity $\sigma_{xy}=\sum_{z}\sigma_{xy}(z)\in\mathbb{Z}$ must fluctuate from sample to sample and cannot be zero for all realizations $\{I\}$. Since states with different total $\sigma_{xy}$ cannot be  deformed to each other without closing the gap, we conclude that the above deformation process must close the energy gap for some disorder realization. This establishes the nontriviality of the mirror SPT state.

\subsection{Brief comments on general cases}
\label{sec:generaldecor}

If the exact and average symmetries form nontrivial group extensions Eq.~\eqref{eq:fullsymmetry}, then we do not have Künneth formula and the decorated domain wall construction will in general become more complicated.

Let us first review the idea of decorated domain walls in clean systems in the general cases (with possibly nontrivial group extensions). In $\mathrm{D}$-spatial dimensions, the construction starts with a phase in which the $G$ symmetry is broken spontaneously. Such a phase admits domain wall excitations, such that a domain wall labeled by an element $g\in G$ interpolates between two symmetry breaking patterns related by a $g$ action. A $G$-symmetric state can be obtained by quantum disordering the symmetry breaking phase, i.e. condensing $G$-domain walls. It is known that the domain wall condensation may give rise to a non-trivial $\mathcal{G}$-SPT phase, if one decorates a $G$-defect in the symmetry breaking phase, i.e. a domain wall or a (multi-)domain wall junction, of codimension $p$ with a $(D-p)$-dimensional SPT phase protected by the unbroken $K$ symmetry \cite{chenlu2014}. Crucially, in order for the condensation of $G$-defects to be SRE, the following consistency conditions must be satisfied:
\begin{enumerate}
    \item $G$-defects of each codimension should be free of $K$-anomaly. Namely, the defects can be gapped without breaking $K$;
    \item $K$ is preserved during a continuous deformation of the $G$-defect network; 
    \item There is no Berry phase accumulated after a closed path of continuous deformation.
    \end{enumerate}
Physically, the third condition is required since the many-body wavefunction is single-valued; the $G$-domain walls can be condensed without breaking $K$ once the second consistency condition is respected; and the first condition guarantees the resulting state to be gapped with a unique ground-state. The wave-function of the gapped $\mathcal{G}$ SPT produced is a superposition of all domain wall patterns. These consistency conditions may be formulated mathematically by the Atiyah-Hirzebruch spectral sequence (AHSS). We refer the reader to \cite{wang2021,2019GaiottoJF} for details. In the decorated domain wall scheme the protected surface states appear naturally: topological defects that end at the surface carry the non-trivial boundary modes of the lower dimensional SPT phases protected by the symmetry $K$.

Now we make $G$ an average symmetry and decorate nontrivial invertible states on $G$ domain walls. The first two conditions above should still be satisfied, since we are interested in SRE ensembles (Def.~\ref{def:SREensemble}). The third condition, however, does not seem to be necessary, since the domain walls no longer form coherent superpositions. This leaves the possibility of \textit{intrinsically disordered} average SPTs that have no counter parts in clean systems. This intriguing possibility is beyond the scope of this work, and will be reported in a subsequent study.

\section{Average anomalies and boundary properties}

\label{sec:boundary}
 
For ordinary SPT, it is well known that a nontrivial bulk leads to nontrivial boundaries. Specifically, the boundary theory will have t'Hooft anomaly that matches the bulk topological response. The t'Hooft anomaly imposes powerful constraints on the IR boundary dynamics. For example, the anomalous boundary cannot be symmetrically gapped with a unique ground state. A natural question is: how does a nontrivial bulk average SPT phase constrain its boundary dynamics? Or equivalently, what are the consequences of an ``average anomaly''? 

As we will see below, the answer to the above question depend on the dimensions of the decorated states on the proliferated domain walls. There are two different categories that we shall discuss separately.

\subsection{The trivial case: $(0+1)d$ decoration}
\label{sec:boringboundary}

Let us illustrate the physics with a simple example. We start from the $(1+1)d$ cluster model \cite{suzuki1971relationship,Son2011}:
\begin{equation}
    H_\mathrm{cluster} =-\sum_n Z_{n-1} X_n Z_{n+1},
    \label{eq:cluster}
\end{equation}
in which $X$ and $Z$ are Pauli matrices. The cluster chain is in an SPT phase protected by a $\mathbb{Z}_2\times \mathbb{Z}_2$ symmetry, which is generated by
\begin{equation}
    K=\prod_{n} X_{2n+1},\quad G=\prod_{n} X_{2n}.
\end{equation}
We then add to the Hamiltonian in Eq.~(\ref{eq:cluster}) disorder that violates one of the $\mathbb{Z}_2$ symmetries, say $G$, but restores it on average. For example, we add the following term
\begin{equation}
    H_{\mathrm{dis}}=-\sum_n h_{2n} Z_{2n},
    \label{eq:disorderHamiltonian}
\end{equation}
where $h_{2n}$'s are onsite potentials distributed uniformly in $[-\delta,\delta]$. The disorder Hamiltonian is symmetric under $K$, while respects $G$ only on average. 

The cluster chain Eq.~\eqref{eq:cluster} can be interpreted as decorating a nontrivial $K$ charge at each $G$ domain wall, and then condense the domain walls to get a $\mathbb{Z}_2\times \mathbb{Z}_2$ symmetric topological phase. Once the random field is turned on, the $G$ domain walls no longer condense as the $G$ symmetry is explicitly broken for each disorder realization. However, for each realization, there will in general be many $G$ domain walls, and each domain wall still traps a nontrivial $K$ charge. The resulting state is therefore a nontrivial $\mathbb{Z}_2\times\mathbb{Z}_2^{(ave)}$ SPT. 

We can in fact push our model to strong disorder regime, and obtain a much simpler effective model:
\beq
H=-\sum_n\left(Z_{2n}X_{2n+1}Z_{2n+2}+h_{2n}Z_{2n}\right),
\eeq
where $h_{2n}\in\{\pm1\}$ are independent binary random variables defined on each even-integer site. The ground state of each individual Hamiltonian is simply an un-entangled product state, with each even site in  $|Z_{2n}=h_{2n}\rangle$ and odd site in $|X_{2n+1}=h_{2n}h_{2n+2}\rangle$. This ensemble has the same domain wall decoration pattern as the previous model (as can be checked explicitly using Eq.~\eqref{eq:DDW}), and is therefore an equally valide (but much simpler) representation of the $\mathbb{Z}_2\times\mathbb{Z}_2^{(ave)}$ SPT. 

The fact that each disorder realization simply gives an un-entangled product state is true even when the system has boundaries. This immediately means that our ``average cluster chain'' does not have nontrivial boundary state -- unlike the clean cluster model which has a robust ground state degeneracy once put on an open chain. This can also be understood directly from the edge state: each end of the clean cluster chain forms a two-dimensional projective representation of $\mathbb{Z}_2\times\mathbb{Z}_2$, in which the generators of $G$ and $K$ act as anti-commuting Pauli matrices $\sigma_x$ and $\sigma_z$, respectively. Now adding, even only on the boundary, a random $G$-breaking field $h\sigma_z$ will lift the edge degeneracy completely. 

We have demonstrated that the $\mathbb{Z}_2\times\mathbb{Z}_2^{(ave)}$ cluster chain does not have nontrivial boundary dynamics. The boundary, however, does have a notable feature: the $K$-charge is fixed by $\sgn(h)$ which fluctuates from sample to sample. This means that different samples will not be symmetrically and adiabatically connected to each other, violating one of the key assumptions of our SRE ensemble (Def.~\ref{def:SREensemble}). So our SPT is similar to the standard SPTs, in the sence that when the system has boundaries the state cannot stay SRE -- although in the above example it violates the SRE condition in a rather trivial way.

It is straightforward to generalize the above observations to all the average SPT states, in any dimensions, in which only $(0+1)d$ states are decorated on average-symmetry domain walls. Such states can be continuously deformed to a limit where each disorder realizations simply gives a product state, without any interesting boundary dynamics. This aspect is in fact familiar in crystalline SPT phases\cite{Huang2017,Fuji2015}: if we decorate $(0+1)d$ states (for example, some integer $U(1)$ charges) on crystalline defects (such as in each unit cell of translation symmetries), we obtain crystalline SPT phases without nontrivial boundary dynamics -- instead we obtain a variety of atomic-like insulators that are not symmetrically and adiabatically connected to each other.

\subsection{Nontrivial cases: higher-dimensional decoration}
\label{sec:nontrivialboundary}

We now move on to the much more interesting cases with higher dimensional domain wall decorations. We shall employ a modified version of flux-insertion argument commonly used in the study of topological phases. Let us again illustrate with a simple example.

Consider a $(2+1)d$ boson SPT, with the exact symmetry$K= SO(3)$, average symmetry $G=\mathbb{Z}_2$ and full symmetry $\mathcal{G}=SO(3)\times \mathbb{Z}_2$. The only non-trivial element in $H^1[G,H^2(K,U(1))]$ has a topological action
\begin{equation}
    S_{\mathrm{top}}=\pi\int_X a\cup w_2^{SO(3)},
    \label{eq:SO(3)Z2}
\end{equation}
where $a$ is the background $\mathbb{Z}_2$ field and $w_2^{SO(3)}$ is the Second Stiefel-Whitney class of the $SO(3)$ probe field. This state has a simple physical picture in terms of decorated domain walls: on each $\mathbb{Z}_2$ domain wall there is a Haldane chain protected by the $SO(3)$ symmetry. 

Let us now put the system on a space manifold with boundary, and ask how likely it is for the ground state $|\Psi\rangle$, for one realization of the disorder potential $v$, to be short-range entangled. We argue below that such ``uninteresting'' ground state must be very rare as the system size becomes large. The trick is to use the partial symmetry transform to create domain walls, similar to the argument used in Sec.~\ref{sec:domain}, but now with a spacial boundary.

\begin{figure}
\begin{center}
  \includegraphics[width=.45\textwidth]{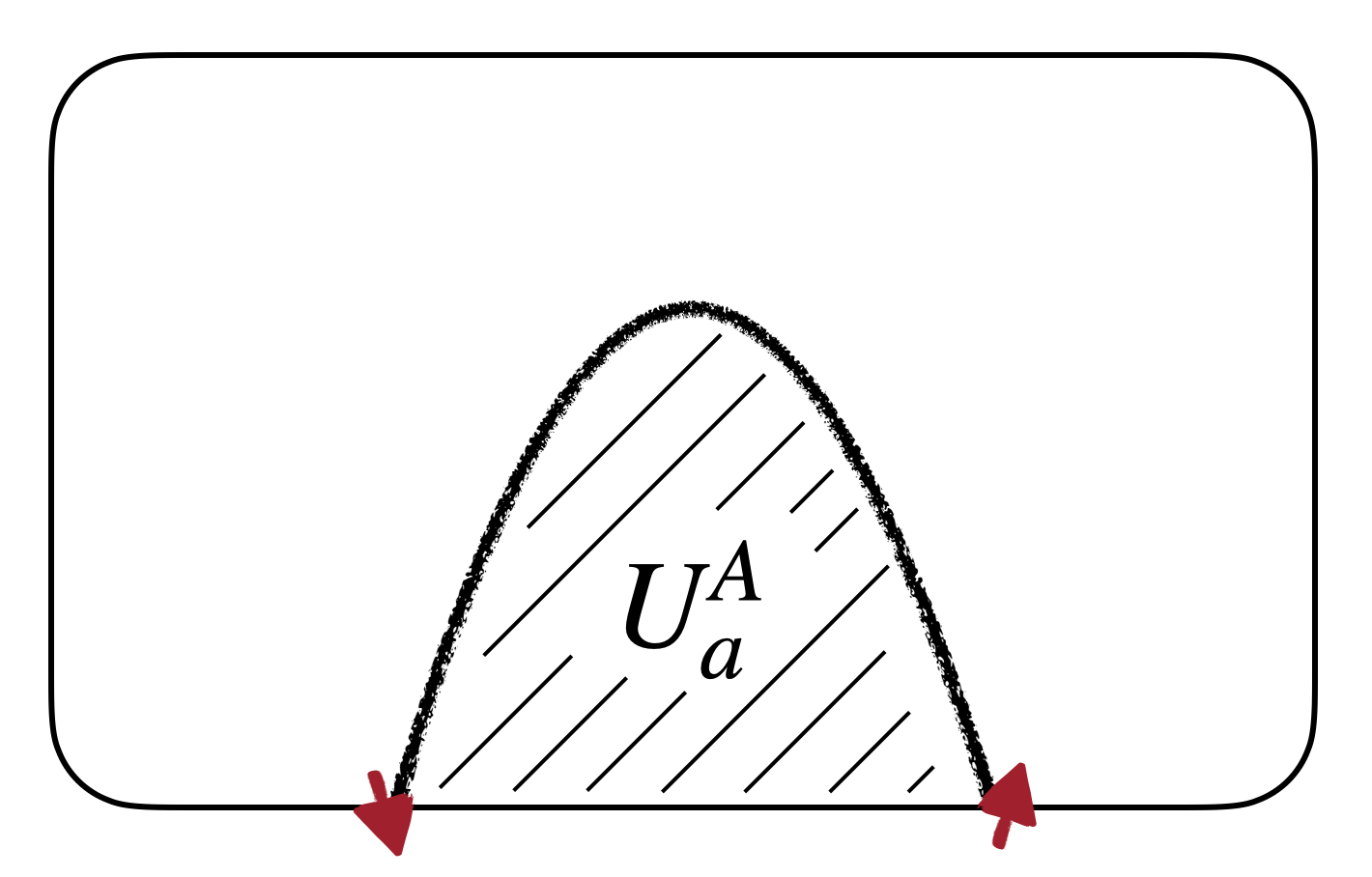} 
\end{center}
\caption{
The decorated domain wall picture in the presence of a physical edge.
}
\label{Fig:Boundary}
\end{figure}

Let us start by assuming that $|\Psi\rangle$ (under a particular $\mathbb{Z}_2$-breaking disorder realization $v$) is short range entangled, with exponentially decaying connected correlation functions and a nonzero energy gap. Now take a large enough sub-region $A$ that includes a segment on the physical edge (Fig.~\ref{Fig:Boundary}), and flip all the random $\mathbb{Z}_2$-breaking fields $v$ inside $A$, so that we are now considering a different disorder realization with
\beq
\tilde{v}(x\in A)=-v(x\in A), \hspace{5pt} \tilde{v}(x\in\bar{A})=v(x\in\bar{A}).
\eeq
We denote the ground state under $\tilde{v}$ as $|\tilde{\Psi}\rangle$. Similar to the bulk argument (Eq.~\eqref{eq:DDW}), we expect that
\beq
U^A_a|\Psi\rangle=V_a^{\partial A}|\tilde{\Psi}\rangle,
\label{eq:DDWboundary}
\eeq
with $V_a^{\partial A}$ creating an $SO(3)$-protected Haldane chain on domain wall $\partial A$ (not including the segment on the physical edge). But contrary to the bulk argument in Sec.~\ref{sec:domain}, the domain wall $\partial{A}$ itself has boundaries -- it terminates on the physical edge at two points. If $|\tilde{\Psi}\rangle$ is also short-range entangled (with correlation length much shorter than the edge segment), then $V_a^{\partial A}$ will create a pair of half-integer spins at the two ends of $\partial{A}$. Since we assume $SO(3)$ to be exactly preserved, the two spins should be locked into a singlet, which leads to a nontrivial correlation at large distance -- the state effectively becomes long-range entangled. But this should not happen, as the left hand side of Eq.~\eqref{eq:DDWboundary} is clearly short-range entangled: it is just a depth-$1$ unitary, $U_a^A$, acting on a short-range entangled state $|\Psi\rangle$. Therefore the assumption that $|\tilde{\Psi}\rangle$ is short-range entangled must be wrong. To make Eq.~\eqref{eq:DDWboundary} valid, $|\tilde{\Psi}\rangle$ must already have a singlet pair distributed at the two ends of $\partial{A}$, so that acting on it with the Haldane chain creation operator $V_a^{\partial A}$ removes the singlet pair and recovers a short-range correlated state.  

Once we understand the long-range correlated (or entangled) nature of $|\tilde{\Psi}\rangle$, it is obvious that such states can be created in many other ways: we can change the region $A$ so that $\partial{A}$ end at different point on the physical edge, we can also have multiple such regions that lead to many long-range singlets on the edge. Crucially, all such states appear with same probability as $|\Psi\rangle$, since, by definition of the average $\mathbb{Z}_2$ symmetry, flipping the sign of the random potential $v$ in a region larger than correlation length should not change its realization probability. Therefore as the system size goes to infinity, there are infinitely many ways to create long-range entanglement out of a short-range entangled state, with essentially equal probability. This in turn means that a short-range entangled state $|\Psi\rangle$ can appear at most with a vanishing probability.

The above argument generalizes to other average SPT phases, as long as the nontrivial invertible states being decorated on the domain walls (defects) are higher than $(0+1)d$. In other words,
\begin{theorem}
An average SPT with decoration dimension $p>(0+1)$ will have long-range entangled boundary state with probability approaching $1$ in the thermodynamic limit.
\label{thm:boundaryanomaly}
\end{theorem}

{ {We emphasize that the above statement does not require ensemble averaging: even for a single sample of disorder realization (which is what we have in real experiments), the boundary theory will be long-range entangled in the thermodynamic limit. Our result also indicates that even with a single sample, in the thermodynamic limit the boundary will have gapless, delocalized excitations. The delocalized gapless excitations will contribute to various measurable quantities such as thermal conductance. Although a detailed account of the dynamical features of the boundary theory may be complicated and require a case-by-case study, we expect the boundary thermal conductance to scale as a power law in temperature $T$ even in the strong disorder regime. Furthermore the nontrivial thermal conductance will disappear once a symmetry-breaking field is turned on, which in principle makes the signal distinguishable from phonon contributions. Such phenomenon could serve as a practical way to experimentally detect nontrivial boundary states with average anomaly.

The nontrivial boundary state, even within a single disorder realization, suggests that the bulk SPT phase should also be well-defined and nontrivial for a single disorder realization. However at this point we do not have a theoretically controlled way to define or describe such ASPT phases with single disorder realization. This is an interesting question for future investigation.
}}

In Table~\ref{classtable} we list, in parenthesis, those states that do have long-range entangled boundary states (with probability approaching unity).

\subsection{Application: Lieb-Schultz-Mattis constraints with average lattice symmetries}
\label{sec:LSM}

Readers familiar with random spin chains will recognize the long-range entangled state constructed in Sec.~\ref{sec:nontrivialboundary} as essentially the random singlet state\cite{ma1979random,dasgupta1980low,fisher1994random}. Indeed, without any change in the argument, we can replace the average $\mathbb{Z}_2$ symmetry in the example of Sec.~\ref{sec:nontrivialboundary} with a $\mathbb{Z}$ symmetry. By the spirit of ``crystalline equivalence principle''\cite{Thorngren2016} we can interpret this $\mathbb{Z}$ as lattice translation. The corresponding bulk system is a stack of $SO(3)$ Haldane chains with an average translation symmetry perpendicular to the chains. On the boundary we obtain a disordered spin-$1/2$ chain with average translation symmetry. The result of Sec.~\ref{sec:nontrivialboundary} then becomes a disordered version\cite{Kimchi2018} of the Lieb-Schultz-Mattis (LSM) theorem\cite{lieb1961two}, which states that a disordered spin-$1/2$ chain with average translation symmetry must stay long-range entangled with probability one. The random singlet state with arbitrarily long-ranged singlet pairs is a classic example of such states.

Using the crystalline equivalence principle\cite{Thorngren2016}, we can conclude that all the generalized LSM anomalies for other lattice symmetries\cite{Po2017,Ye2021} (rotation, reflection etc.) still imply long-range entanglement (with probability $1$) when the lattice symmetry becomes average.

Let us provide a more direct and detailed argument for the simple case of $(1+1)d$ systems with average lattice translation symmetry. Consider a spin chain with exact on-site symmetry $K$, with the Hilbert space for each lattice unit cell forming a projective representation $\omega_{uc}\in H^2(K,U(1))$. For concreteness we can think of $K=SO(3)$ and the system being a spin-$1/2$ chain, although this will not be necessary. 

Now assume that for some disorder realization (with a local Hamiltonian $H=\sum_i H_i$), the ground state $|\Psi\rangle$ is short-range entangled with a finite correlation length $\xi$. Let us then consider a different Hamiltonian $\tilde{H}=\sum_i \tilde{H}_i$, defined with a large subregion (a long segment) $A$, such that
\begin{enumerate}
    \item for $i$ far outside $A$, $\tilde{H}_i=H_i$,
    \item for $i$ deep inside $A$, $\tilde{H}_i=H_{i-1}$,
    \item for $i$ near the boundary $\partial A$, $\tilde{H}_i$ can take any value in the ensemble.
\end{enumerate}
Essentially we have translated the Hamiltonian inside region $A$ by one unit cell, which is the translation analogue of the partial symmetry operation in Sec.~\ref{sec:nontrivialboundary}. This disorder realization will have a different probability with $H$, but crucially the two probabilities only differ by a constant factor, depending on details at $\partial A$ but independent of either the size or location of region $A$ (as long as $A$ is large enough). 

Since we have assumed the original state $|\Psi\rangle$ to be short-range correlated with a clear energy gap, the change in a local Hamiltonian term (say at $i$) should only affect properties near $i$. So the new ground state $|\tilde{\Psi}\rangle$ should be identical to $|\Psi\rangle$ far out of $A$, and be identical to the translated version $T_x|\Psi\rangle$ deep inside $A$. However, these two conditions imply that at each boundary $\partial A$ there is an extra half-integer spin (or projective representation in general). In order to form a symmetric state, these two half-integer spins have no choice but to form a singlet with each other (since regions deep inside and far outside of $A$ are determined already). This creates a long-range correlation across the large region $A$.

Let us make the above argument more explicit in terms of reduced density matrices. We denote a sub-segment deep inside $A$ as $A_-$, the region far outside $A$ as $\overline{A_+}$, and the remaining two regions (the left and right boundaries) as $\partial A_L$ and $\partial A_R$. We further denote $\widetilde{A_-}$ as $A_-$ translated to the right by one unit cell, $\widetilde{\partial A}_L$ as $\partial A_L$ plus one unit cell right to it, and $\widetilde{\partial A}_R$ as $\partial A_R$ minus its leftmost unit cell. We now consider reduced density matrices from the state $|\Psi\rangle$ (denoted as $\rho$) and from the state $|\tilde{\Psi}\rangle$ (denoted as $\tilde{\rho}$. For an SRE state, at each of the four entanglement cuts (let us denote as $a,b,c,d$ from left to right) we can extract an element of $\omega\in H^2(K,U(1))$ from the entanglement spectrum\cite{Pollmann2010} (for $K=SO(3)$ this $\mathbb{Z}_2$ number is just measuring the parity of singlet bonds across each cut). Since we have a nontrivial $\omega_{uc}\in H^2(K,SO(3))$ per unit cell, we have the relations $\omega_a-\omega_b=\omega_{uc}\times |\partial A_L|$ and $\omega_c-\omega_d=\omega_{uc}\times |\partial A_R|$. Now the SRE nature of $|\Psi\rangle$ and the relation between $\tilde{H}$ and $H$ imply that $\rho(\overline{A_+})=\tilde{\rho}(\overline{A_+})$ and $\rho(A_-)=\tilde{\rho}(\widetilde{A_-})$.  Therefore at each of the four entanglement cuts we should have $\omega=\tilde{\omega}$ (now $\tilde{b}$ and $\tilde{c}$ are translated from $b$ and $c$ by one unit cell). However, this means that for the two boundary regions, $\tilde{\omega}_a-\tilde{\omega}_b=\omega_{uc}\times (|\widetilde{\partial A}_L|-1)$ and $\tilde{\omega}_c-\tilde{\omega}_d=\omega_{uc}\times (|\widetilde{\partial A}_R|+1)$. Therefore the two regions $\widetilde{\partial A}_L\cup \widetilde{\partial A}_R$ cannot be short-range entangled -- the only way to have a symmetric state is for the two regions, separated by $\widetilde{A_-}$, to entangle with each other.

We can now make the above argument for any large region $A$, even multiple of them. Since the probability to create such long-range correlation does not depend on the size and location of $A$, we again conclude that for such systems, short-range entangled ground state must be extremely rare, with at most vanishing probability as system size $L\to \infty$.

We note that for $K=SO(3)$, a similar average LSM theorem have been shown in Ref.~\cite{Kimchi2018}. Our argument here is more general, although the conclusion is not as strong -- for example, we make no direct statement about averaged correlation functions or energy gaps.

\section{Fermionic examples}
\label{sec:fermionicexample}

{ 
The insight we obtained in Sec.~\ref{sec:domain} works equally well for systems with fermions and/or beyond the group cohomology classification.
In this section, we first discuss some known examples of nontrivial free fermion ASPT states from previous literature. As a demonstration of the power of our approach, we will reproduce several nontrivial results in a straightforward manner. We will then systematically discuss two particularly interesting symmetry classes of fermionic ASPT phases, namely $(3+1)d$ fermionic TIs in symmetry class AII and AIII. We study the former using a systematic decorated domain wall construction similar to that in Ref.~\cite{Song2017}, and the latter by examining the reduction of the clean classification.  
}

\subsection{ {Known examples}}
\label{sec:known}

\textbf{3D TI}: the $(3+1)d$ Fu-Kane-Mele topological insulator\cite{Fu2007}, protected by charge $U(1)$ and Kramers time-reversal symmetry, can be viewed as decorating $(2+1)d$ time-reversal domain walls in the bulk with integer quantum Hall states with Hall conductance $\sigma_{xy}=1$ (mod $2$). Since the decorating dimension is $p=2>0$, the state will remain nontrivial, with nontrivial surfact states, as we break time-reversal to an average symmetry (the total symmetry being $U(1)\rtimes(\mathbb{Z}_2^T)^{ave}$). This is in agreement with Ref.~\cite{Fu2012}, where it was found that the TI surface remain delocalized even in the presence of magnetic impurities.

\textbf{3D weak TI}: the $(3+1)d$ weak topological insulator\cite{Fu2007}, protected jointly by charge $U(1)$, Kramers time-reversal and translation symmetry, can be viewed as a stack of $(2+1)d$ Kane-Mele topological insulator\cite{Kane2005} (protected by $U(1)\rtimes \mathbb{Z}_2^T$) in one spacial direciton (call it $\hat{z}$). The layers being stacked can be viewed as $(2+1)d$ defect of the translation symmetry. So the docoration dimension here is $p=2>0$. This means that if we break translation symmetry down to an average symmetry while keeping $U(1)\rtimes\mathbb{Z}_2^T$ exact (a very natural condition for realistic crystals), the state will remain nontrivial with nontrivial surface states (the total symmetry being $(U(1)\rtimes \mathbb{Z}_2^T)\times \mathbb{Z}^{ave}$). This agrees with Refs.~\cite{Ringel2012,Mong2012}, where it was found that the surface theory remains delocalized even with average translation symmetry.

\textbf{2D TI}: the $(2+1)d$ Kane-Mele topological insulator\cite{Kane2005}, protected by charge $U(1)$ and Kramers time-reversal symmetry, can be constructed\cite{Lan2019} via decorating $(0+1)d$ time-reversal defects (intersections of domain walls) with odd-integer $U(1)$ charge. Since the decoration dimension is $p=0$, as we break time-reversal down to an average symmetry, even though the bulk is still considered nontrivial in our definition, the edge state can be trivialized (the total symmetry being $U(1)\rtimes(\mathbb{Z}_2^T)^{ave}$). This agrees with Refs.~\cite{Altshuler2013,Chou2dTI}, where it was found that the helical edge state of the Kane-Mele TI can become localized when time-reversal symmetry is broken (spontaneously or explicitly) to an average symmetry.

\subsection{Class AII}
\label{sec:AII}

Let us systematically consider 3D TIs protected by $U(1)\rtimes \mathcal{T}$ symmetry (class AII), in which $U(1)$ is the electron charge conservation and $\mathcal{T}$ is time reversal, with $\mathcal{T}^2=-1$ when acting on fermionic operators. Importantly, $\mathcal{T}$ preserves the $U(1)$ charge. We consider the case where $\mathcal{T}$ becomes an average symmetry, while charge conservation remains exact. 
As illustrated in Sec.~\ref{sec:domain}, the ASPT phases in our symmetry setting can be constructed by decorating a $(3-p)$D fermionic $U(1)$ SPT on each codimension-$p$ (with respect to the 3D space) $\mathcal{T}$-symmetry defect. The first two consistency conditions listed in Sec.~\ref{sec:generaldecor} need to be satisfied. In the disorder setting, the first condition ensures each state in the mixed ensemble is $U(1)$-symmetric and SRE, while the second guarantees any pair of states can be adiabatically connected without breaking $U(1)$ -- this is precisely our definition for a $U(1)$ symmetric SRE ensemble. Specifically, the construction follows the guideline below:
\begin{itemize}
    \item One starts from the top codimension $p=0$, and decorates $\mathcal{T}$-defects of increasing $p$ successively;
    \item The quantum anomalies must cancel out on codimension $p$ defects, given all previous decorations with codimensions $p'<p$;  
    \item After the $p=3$ decoration, the second consistency condition in Sec.~\ref{sec:domain}, i.e. the constraints on continuous deformations of domain walls, must be satisfied.
\end{itemize}
In this section we present the decorated defect construction in a physical way. A rigorous AHSS calculation can be found in Appendix.~\ref{app:SSS}.

Let us start with codimension-0 defects, namely, the 3D patches in which $\mathcal{T}$ is broken by the disorder. It is known that fermionic SPT phases protected by $U(1)$ symmetry are classified by the $\mathrm{spin}^c$ cobordism group of a point, $\Omega_{\mathrm{spin}^c}^{\bullet}(\mathrm{pt})$ \cite{Kapustin2015,Freed2016,Guo2018,Garc2019}. In particular, there is no 3D non-trivial phase protected by $U(1)$ alone. {Therefore, all 3D patches are in the trivial $U(1)$ symmetric SRE phase.} 

We then move on to codimension-1 $\mathcal{T}-$domain walls between the patches. Since two adjacent patches are both in the same (trivial) phase, the wall in between traps no anomalous surface mode and can thus be gapped without breaking the $U(1)$ symmetry. One now decorates the $\mathcal{T}$-domain wall with 2D $U(1)$ SPT phases. There are two non-trivial choices: the integer quantum Hall (IQH) state and the Kitaev $E_8$ state \cite{kitaev2011toward}, each of which has a $\mathbb{Z}$ classification. We label the two integers by $n_I$ and $n_E$ for the IQH state and the $E_8$ state respectively. The elementary $E_8$ state with $n_E=1$ has 8 chiral bosons at the edge, which can be thought of as protected by a gravitational anomaly, whose “probe field" is the background space-time geometry.

Naively, one may expect decorating 2D layers labeled by different integers leads to different 3D SPT phases. However, this is not the case. The easiest way to see this is to consider decorating an IQH state with $n_I=2$ on the $\mathcal{T}$-domain wall. When a domain wall is cut open at the surface of the system, a helical edge state with chiral central charge $c=2$ appears. We can deposit $n_I=\pm1$ IQH states on the surfaces of the domains, such that at the surface $\mathcal{T}$-domain wall boundary there arises chiral modes with $c=-2$. {The two counter-propagating modes can be trivialized by turning on a coupling, resulting in a unique gapped ground-state both in the bulk and on the surface\footnote{ As argued in Sec.~\ref{sec:boundary}, if the surface can be made SRE in the presence of a bulk decoration with dimension $p>(0+1)$, this decoration is guaranteed to be trivial.  }. The same argument also applies to the $E_8$ decoration, which implies that the indices $n_I$ and $n_E$ are only defined modulo 2. This argument resembles the operation of adjoining layers in Ref.~\cite{Song2017}. In summary, for $p=1$ we have two possible decorations, each of which is labeled by $\mathbb{Z}_2$.\footnote{Mathematically, each of the two decorations is described by the cohomology $H^1(\mathbb{Z}_2^\mathcal{T},\mathbb{Z}^\mathcal{T})=\mathbb{Z}_2$, where $\mathbb{Z}^\mathcal{T}$ denotes the twisted coefficient, reflecting the fact that time reversal acts non-trivially on the IQH and $E_8$ states.}}
One more comment is hereafter we require any two defects that can be smoothly deformed into each other to be decorated by the same lower dimensional phase. This is due to the assumption that states in different disorder realizations should be adiabatically connected (Def.~\ref{def:SREensemble}).

\begin{figure}
\begin{center}
\begin{subfigure}[b]{0.2\textwidth}
  \includegraphics[width=\textwidth]{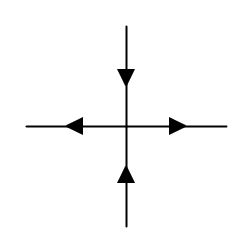} 
  \caption{\label{fig:codim2defect}}
  \end{subfigure}
  \begin{subfigure}[b]{0.2\textwidth}
  \includegraphics[width=\textwidth]{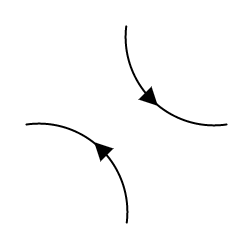} \caption{\label{fig:codim2defectresolved}}
  \end{subfigure}
\end{center}
\caption{
(a) An intersection of two $\mathcal{T}$-domain walls, which is decorated by an IQH/$E_8$ state, viewed from the top. Solid lines represent an IQH/$E_8$ state on each half plane, with chiralities of edge modes indicated by the arrows.
(b) To see there is no gapless chiral mode at the intersection, note that it can be smoothly deformed into two disjoint walls.
}
\label{Fig:codim2}
\end{figure}

Next we proceed to codimension-2 $\mathcal{T}$-defects, i.e. the 1D intersections of $\mathcal{T}$-domain walls. { We should first examine whether the possible decorations in lower codimensions lead to any quantum anomaly. A 1D domain wall intersection is shown in Fig.~\ref{fig:codim2defect}, with an IQH/$\mathrm{E}_8$ decorated on each domain wall. (Remember that the edge chirality is defined only mod 2.) The intersection has no net chirality and can thus be gapped without $U(1)$ symmetry breaking.\footnote{An intersection of $\mathcal{T}$-domain walls is trivial, in the sense that it can be deformed locally to the configuration in Fig.~\ref{fig:codim2defectresolved}. The two configurations differ at most by a 1D SRE state. This observation leads to the same result that the domain wall intersection traps no 1D gapless mode.} 
We then consider decorating a domain wall intersection with 1D fermionic SPT phases protected by the exact $U(1)$ symmetry. However, there is no non-trivial 1D SPT protected by $U(1)$ alone. Therefore we do not have any new decoration at codimension 2.}

One can repeat the same procedure for codimension-3 $\mathcal{T}$-defects, namely, 0D points, each of which is an intersection of three domain walls. {It is straightforward to see that a 0D defect can always be gapped without breaking the exact $U(1)$ symmetry, given all the previous decorations. The reason is simply that there is no non-trivial $U(1)$ SPT in 1D, whose anomaly inflow can protect a zero mode in 0D as a boundary state. As a result, for each \emph{quenched} realization of disorder pattern, $\mathcal{T}$-symmetry defects in all codimensions can be trivially gapped out.

By assumption in Def.~\ref{def:SREensemble}, we demand that the $U(1)$ charge must be conserved when the $\mathcal{T}$ domain walls are deformed continuously. The decorations by the IQH and the $E_8$ state are consistent with this constraint, as shown by an explicit spectral sequence calculation in Appendix.~\ref{app:SSS}. On the other hand, one may decorate each 0D $\mathcal{T}$-defect point using 0D $U(1)$ SPT states. These $U(1)$ SPT states have a $\mathbb{Z}$ classification, whose physical meaning is the $U(1)$ charge quantum number carried by the 0D ground-state. For symmetry class AII, a $\mathcal{T}$ transformation preserves the $U(1)$ charge. The only decoration consistent with the fusion rule of 0D $\mathcal{T}$-defects is the trivial one 
(Note that 0D point-like $\mathcal{T}$-defects can annihilate in pairs, thus decorating charges on them is forbidden since it breaks the $U(1)$ charge conservation during the deformation of $\mathcal{T}$ defects.), which is described by the cohomology $H^3(\mathbb{Z}_2,\mathbb{Z})=0$.\footnote{Here the coefficient is untwisted, as $\mathcal{T}$ preserves the $U(1)$ charge in class AII.} As a result, there is no new possible decoration at 0D (codimension 3).}

{At this point, we have exhausted all possible decorations on $\mathcal{T}-$defects of all codimensions, and have also ensured the consistency, i.e. the domain wall condensation has a unique gapped ground-state with the decorations described. 
We thus reach our final result: when $\mathcal{T}$ is restored on average, 3D TIs in symmetry class AII are classified by $\mathbb{Z}_2^2$}, generated by placing an IQH state or an $E_8$ state on the $\mathcal{T}$ domain wall, respectively. Moreover, since both decorations are extended in space (2D), from Theorem~\ref{thm:boundaryanomaly} we conclude that all the non-trivial ASPTs in this symmetry class have long range entanglement on the surface with probability one in the thermodynamic limit.

The classification of clean $3D$ TIs in class AII is $\mathbb{Z}_2^3$\cite{WangPotter2014,Wang2014,Freed2016}. In comparison, our $\mathbb{Z}_2^2$ classification for the ASPT in this symmetry class misses one nontrivial state. The missing state, known as $eTmT$ state, can be obtained from the domain wall condensation approach with some nontrivial phase factors in the domain wall condensate. As we explained in Sec.~\ref{sec:domain}, such state is no longer nontrivial when the symmetry becomes average, as coherent superpositions are replaced by classical probabilities, and the notion of superposition phase factors is no longer well defined. From the topological response point of view, the topological effective action of the $eTmT$ state reads \cite{Kapustin2014boson}
\begin{equation}
    S_{\mathrm{top}}=\pi\int_X w_1^4,
    \label{eq:eTmT}
\end{equation}
where $w_1$ is the first Stiefel-Whitney class (``time reversal gauge field'') of the worldvolume of the $(3+1)d$ bulk. The average nature of time-reversal gives the constraint
\begin{equation}
    \int_\tau w_1=0,
\end{equation}
under which the TQFT in Eq.~(\ref{eq:eTmT}) vanishes identically.

\subsection{Class AIII}

We now study the disorder classification of 3D TIs with symmetry group $U(1)\times \mathcal{T}$ (class AIII), with $\mathcal{T}$, sometimes also called ``particle-hole symmetry'' as in quantum Hall context, being an average symmetry. Unlike the electric charge, now the $U(1)$ charge is odd under time reversal. 

For simplicity, here we explicitly focus on the clean SPT phases, classified by $\mathbb{Z}_8\times \mathbb{Z}_2$ \cite{Wang2014}, and ask which of these phases remain nontrivial as $\mathcal{T}$ becomes an average symmetry. The $\mathbb{Z}_2$ factor corresponds to the $e_fm_f$ state, which as we showed in Sec.~\ref{sec:AII} remains non-trivial in the presence of disorder. Similarly, the $n=1$ state in the $\mathbb{Z}_8$ can be understood as decorating an IQHE on the $\mathcal{T}$-domain walls, which remains nontrivial as argued also in Sec.~\ref{sec:AII}. The $n=4$ state in the $\mathbb{Z}_8$ factor is known to be equivalent to the bosonic $eTmT$ state, so from our argument in Sec.~\ref{sec:AII} it should become trivial once $\mathcal{T}$ becomes average. The only nontrivial question now is what happens to the $n=2$ state.

In the clean setup, this state can be constructed  by decorating a unit $U(1)$ charge at each $0D$ intersection of three $\mathcal{T}$-domain walls\footnote{This is only allowed by the defect fusion rule when $\mathcal{T}$ reverses the $U(1)$ charge, which is the case for class AIII.}. 
This is a nontrivial decoration pattern, as    
the $U(1)$ charge decorated at each $0D$ $\mathcal{T}$-defect can not be removed as long as $U(1)$ remains exact.\footnote{Mathematically, 3D TIs in class AIII is classified by the cobordism group $\Omega_{\mathrm{pin}^c}^4(\mathrm{pt})=\mathbb{Z}_8\times \mathbb{Z}_2$, which is an iterated extension of
\begin{align}
    & H^1(\mathbb{Z}_2^\mathcal{T},\mathbb{Z}^\mathcal{T}\oplus \mathbb{Z}^\mathcal{T})=\mathbb{Z}_2 \times \mathbb{Z}_2 \\
    \mathrm{by} \, & H^3(\mathbb{Z}_2^\mathcal{T},\mathbb{Z}^\mathcal{T})=\mathbb{Z}_2 \\
    \mathrm{by} \, & H^5(\mathbb{Z}_2^\mathcal{T},\mathbb{Z}^\mathcal{T})=\mathbb{Z}_2. 
\end{align}
The physical meaning is that $n=2$ mod 4 elements in the $\mathbb{Z}_8$ factor have a $U(1)$ charge decorated on each codimension 3 ($0D$) time reversal defect. }
So we conclude that the bulk state should remain nontrivial as $\mathcal{T}$ becomes average symmetry. However, since the $U(1)$ SPT phase decorated on the $\mathcal{T}$ defect is in $0D$ (a charge), there is no protected surface state for the $n=2$ state based on the discussion in Sec.~\ref{sec:boringboundary}.

To summarize, the final classification for 3D TIs in class AIII with average time reversal symmetry is $\mathbb{Z}_4\times \mathbb{Z}_2$, in which the $n=2$ state in the $\mathbb{Z}_4$ factor has no symmetry protected long range entanglement on the surface. 

We make a comment in connection to the (disordered) integer quantum Hall plateau transition. The average particle-hole symmetry, relating filled and empty Landau levels, emerges naturally at the plateau transition. The resulting $U(1)\times \mathcal{T}^{(ave)}$ has the same anomaly as the $n=1$ state in the $\mathbb{Z}_8$ factor (in clean limit). Our result shows that the plateau transition in two-layer systems ($n=2$ in $\mathbb{Z}_8$), even though being technically ``anomalous'', is not protected to be long-range entangled. This is consistent, in a nontrivial manner, with the numerical fact that such transition can indeed be Anderson localized.

\section{Generalized quantum disorder: a quantum channel approach}
\label{sec:generalizeddisorder}

So far we have treated disorder as purely classical degrees of freedom. However, real disorders, such as impurities in solids, are quantum mechanical, and in principle can develop interesting quantum entanglement within themselves (even though these may not be energetically favorable in typical conditions). In this section, we generalize our considerations to disorders that can develop invertible quantum many-body entanglements. This is a minimal quantum mechanical generalization of disorder, as the disorder potential still remain short-range correlated. We dub such disorders \textbf{invertible quantum disorders}. The observables of our interest, however, will still only live in the ``dynamical'' Hilbert space that does not involve the disorder degrees of freedom. In other words, the disorders are traced out, leaving behind a mixed state. This motivates us, in this section, to develop an SPT theory for such mixed state based purely on the density matrix  $\rho=\sum_IP_I|\Psi_I\rangle\langle\Psi_I|$ ($I$ labeling each ``disorder realization'' in the generalized sense), without referring to any parent Hamiltonian. For this purpose, we will first need to modify some notions in Sec.~\ref{sec:general}, including SRE ensembles, exact and average symmetries, so that these notions are defined purely in terms of the density matrix $\rho$.

\subsection{Symmetries and short-range entanglement}
\label{subsec:symmetryconditions}

As mentioned in the Introduction, in clean systems an SPT has a symmetric SRE ground-state, yet which can not be deformed to a trivial product state using a finite depth quantum circuit if certain symmetries are imposed. To be clear on what states one should consider in the presence of invertible disorder, we need a mixed state generalization of SRE state and the symmetry conditions to which 
it is subject.  

Let us consider a discrete lattice $\Lambda$ in $d$ dimensional space. The total Hilbert space $\mathcal{H}$ is a tensor product of local Hilbert spaces placed at each lattice site, $\mathcal{H}=\otimes_{i\in \Lambda} \mathcal{H}_i$. One can define the notion of SRE mixed state, purely based on the density matrix, following Hastings \cite{Hastings2011}:  

\begin{definition}
\label{def:SRErho}
 Let $\rho$ be the density operator of a mixed state, acting on the Hilbert space $\mathcal{H}$. $\rho$ is SRE if it has a SRE purification. Specifically, there exist the following:
\begin{itemize}
    \item An enlarged Hilbert space $\mathcal{H}'=\mathcal{H}\otimes \mathcal{D}$, constructed by tensoring in additional degrees of freedom on each site;
    \item A SRE pure state $|\psi\rangle$ defined in the Hilbert space $\mathcal{H}'$, such that
    \begin{equation}
        ||\rho - \mathrm{tr}_\mathcal{D}(|\psi\rangle \langle \psi |) ||_1 < \epsilon,
        \label{eq:mixedSRE}
    \end{equation}
    with vanishing $\epsilon$ in the thermodynamic limit (the system size $L\to\infty$). Here the $||...||_1$ denotes the trace norm, which for a Hermitian operator is the sum of the absolute values of its eigenvalues.  
\end{itemize}
\end{definition}
Physically, an SRE mixed state is one that can be obtained from an SRE pure state by tracing out ancillas defined locally on each site. In disorder systems, it is instructive to think of the ancillary space $\mathcal{D}$ as describing the disorder and the partial trace of $\mathcal{D}$ as encoding how the system of interest (in the Hilbert space $\mathcal{H}$) is affected by the interaction with disorder. For this Section, we will focus on disorder ensembles that are SRE in the sense of Definition.~\ref{def:SRErho}. We should emphasize that such purification is in general not unique, and we will not focus on properties that are sensitive to details of the SRE purification -- its mere existence is enough for our purpose.

Analogous to the clean case, the density operator $\rho$ and quantum circuits implemented on $\rho$ are subject to some symmetry conditions. For a moment, let us focus on onsite \textit{unitary} symmetries. As before, we consider two distinct types of symmetries in this work. The first is the exact symmetry, intuitively, the symmetry respected by all possible realizations of disorder. We denote the exact symmetry group by $K$. For each element $k\in K$, there is a corresponding unitary operator $U(k)$ acting on $\mathcal{H}$, which forms a linear representation of $K$:
\begin{equation}
    U(k)=\otimes_{i\in \Lambda}u_i(k),
\end{equation}
where $u_i(k)$ is the (linear) representation of $K$ on a single site $i\in \Lambda$. We generalize the concept of symmetric quantum state to mixed ensembles as following. 

\begin{definition} 
An SRE mixed state $\rho$ has an exact unitary symmetry $K$, if there exist
\begin{itemize}
    \item an enlarged Hilbert space $\mathcal{H}'$ with symmetry action
\begin{equation}
    \tilde{S}(k)=U(k)\otimes \mathbf{1}^\mathcal{D};
    \label{eq:exactsym}
\end{equation}
    \item a SRE purification $|\psi \rangle$ of $\rho$, defined in the enlarged space $\mathcal{H}'$, such that $| \psi \rangle$ is an eigenstate of $S(k)$ for each $k\in K$.  
\end{itemize}
\end{definition}

Note that the ancillary Hilbert space $\mathcal{D}$ is in a trivial representation of $K$. It is not difficult to show that, if an SRE $\rho$ has an exact symmetry $K$, it can be decomposed into an incoherent sum of pure states, which are all eigenstates of $U(k)$ with the \emph{same} eigenvalue.

We now define average symmetry $G$ for our mixed state. The hallmark of an average symmetry is that disorders also transform nontrivially. This motivates the following definition:

\begin{definition}
 An SRE mixed state described by a density operator $\rho$ has an average unitary symmetry $G$ if
\begin{itemize}
    \item there exists a SRE purification $|\psi \rangle$ of $\rho$, defined in an enlarged space $\mathcal{H}'$ with symmetry action
    \begin{equation}
    \tilde{S}(g)=U(g)\otimes U(g)^\mathcal{D},
    \label{eq:averagesym}    
    \end{equation}
    such that $| \psi \rangle$ is an eigenstate of $S(g)$ for each element $g$ in group $G$.  
\end{itemize}
\end{definition}

We emphasize that the ancillary space $\mathcal{D}$ is in a \emph{non-trivial} representation of $G$. With this definition, a density matrix $\rho$ with average symmetry $G$ commutes with the operator $U(g)$ (both viewed as operators acting on the Hilbert space of interest $\mathcal{H}$):
\begin{equation}
    \begin{split}
        U(g)\rho=& \mathrm{tr}_\mathcal{D}[(U(g)^\mathcal{D})^\dagger U(g)\otimes U(g)^\mathcal{D} |\psi\rangle \langle \psi |] \\
        =& \mathrm{tr}_\mathcal{D}[(U(g)^\mathcal{D})^\dagger |\psi\rangle \langle \psi | U(g)\otimes U(g)^\mathcal{D}]=\rho U(g),
    \end{split}
\end{equation}
which is consistent with our expectation for a “statistical symmetry" that is respected on average. A key difference from an exact symmetry is that, when we simultaneously diagonalize the density operator $\rho$ and $U(g)$, $\rho$ is written as an incoherent sum of pure states, with in general different charges under $G$.

We are now ready to discuss \emph{relations} between SRE ensembles. In the standard theory of SPT, quantum states are divided into equivalence classes, where two states are in the same phase iff they can be connected by a symmetric finite-depth local unitary. Naturally, for mixed ensembles, the state equivalence relation can be defined using “symmetric finite-depth" quantum channels\cite{deGroot2021arXiv}. In general, a quantum channel, which is a completely positive trace-preserving map between density operators, can be realized by a unitary acting on an extended system \cite{preskill1998lecture}. We therefore define symmetric finite-depth local quantum channels as following.

\begin{definition}
\label{def:SymChannel}
A quantum channel $\mathcal{E}$ on a system with Hilbert space $\mathcal{H}$ is a symmetric finite-depth local quantum channel if it has a purification to a unitary $W$ on a space $\mathcal{H}'' = \mathcal{H}\otimes \mathcal{A}$, such that for some ancilla state $| a \rangle \in \mathcal{A}$,
\begin{equation}
    \mathcal{E}(\rho) = \mathrm{tr}_\mathcal{A}[W (\rho\otimes|a\rangle\langle a|) W^\dagger].
        \label{eq:symmetricchannel}
\end{equation}
Specifically, we have 
\begin{itemize}
    \item The ancillary space $\mathcal{A}$, which is a tensor product of local degrees of freedom at each site, should \emph{not} be confused with the space $\mathcal{D}$ that is used to purify the density operator $\rho$. However, $\mathcal{A}$ carries the \emph{same} symmetry representation as the disorder (and the space $\mathcal{D}$);
    \item $W$ is a finite-depth local unitary on $\mathcal{H}''$;
    \item $W$ is composed of gates that commute with $S(k) = U(k) \otimes \mathbf{1}^\mathcal{A}$ and $S(g) = U(g) \otimes U(g)^\mathcal{A}$, but do not commute with $U(g)$ that acts on $\mathcal{H}$ alone;
    \item The ancilla state $| a \rangle$ is a product state symmetric under $U(g)^\mathcal{A}$.
   
\end{itemize}
\end{definition}

One can easily check that a symmetric quantum channel preserves exact and average unitary symmetries of an ensemble. Physically, this means that when we apply the quantum channel, the mixed ensemble does not exchange $K$ charge with the ancillas in $\mathcal{A}$. On the other hand, the total $G$ charge of $\mathcal{H}$ and $\mathcal{A}$ is conserved, though there can be charge exchange between them.

We now comment on time reversal symmetry $\mathcal{T}$. As time reversal is anti-unitary, there is no way for the ancillary Hilbert space $\mathcal{D}$ to transform trivially like Eq.~\eqref{eq:exactsym}. Meanwhile, one cannot tell whether a mixed state is an exact or average eigenstate by the $\mathcal{T}$-“charges" when written as an incoherent sum, since time-reversal eigenvalue is anyway a basis-dependent quantity. At best we can define a mixed state $\rho$ to be time-reversal invariant when
\begin{equation}
    \mathcal{T} \rho \mathcal{T}^{-1}= \rho.
\end{equation}
An equivalent statement is that $\rho$ has a purification $|\psi\rangle$ defined in an enlarged Hilbert space $\mathcal{H}'$, such that $|\psi\rangle$ is an eigenstate of time reversal symmetry $\mathcal{T}$. We therefore conclude that, with quantum disorders, \textit{time-reversal symmetry always behaves as an average symmetry}.

After introducing the mixed state generalization of SRE states and the definition of symmetric quantum channels, we are now ready to define the concept of average SPT in terms of the density operator $\rho$.

\subsection{Average symmetry-protected topological phases}
\label{subsec:ASPT}

We now propose the following channel definition of Average Symmetry-Protected Topological phases (ASPT) in the precense of invertible quantum disorders.  

\begin{definition}
\label{def:channelSPT}
Consider two SRE ensembles $\rho_1$ and $\rho_2$, with exact symmetry $K$ and average symmetry $G$. 
\begin{itemize}
    \item  $\rho_1$ and $\rho_2$ are in the same ASPT phase if there exist two symmetric finite-depth local quantum channels $\mathcal{E}$ and $\mathcal{E}'$, such that both $||\mathcal{E}(\rho_1)-\rho_2||_1$ and $||\mathcal{E}'(\rho_2)-\rho_1||_1$ vanish in the thermodynamic limit;
    
    \item In particular, a symmetric SRE $\rho$ is a trivial ASPT if it is two-way connectable to a product state. Namely, there exist two symmetric finite-depth local quantum channels $\mathcal{E}$ and $\mathcal{E}'$, such that 
\begin{equation}
\begin{split}
    \lim_{L\to \infty}|| \rho - \mathcal{E}(\rho_{cl}) ||_1 \to 0,\\
    \lim_{L\to \infty}|| \rho_{cl} - \mathcal{E}'(\rho) ||_1 \to 0.
\end{split}
\label{eq:ASPTdefinition}
\end{equation}
Here the density operator $\rho_{cl}$ represents a pure symmetric product state in the Hilbert space $\mathcal{H}$ and $L$ is the linear size of the system.
\end{itemize}
\end{definition}

Several comments follow. (1) An SPT phase in a clean setting is an eigenstate of the protecting symmetry. As an analog, an ASPT is a mixed ensemble symmetric under the pertinent exact (average) symmetries. This property is preserved by symmetry finite-depth local quantum channels. (2) Quantum channels are generically not invertible, and form a semigroup under composition. Consequently, the above definition for ASPT is an equivalence relation, according to which states are divided into equivalence classes (phases). The physical idea is that two SRE mixed states are in the same ASPT phase if we can prepare each one from the other, using a symmetric finite-depth local channel (potentially with ancillas). In particular, an SRE ensemble is trivial when it can be prepared in this way starting from a trivial product state. (3) When constructing the symmetric finite-depth local channel, the maximal width of the gates is bounded by some constant. The depth of a channel is allowed to be PolyLog($L$) to simulate an adiabatic evolution more accurately \cite{Osborne2007,Coser2018,Haah2018}. However, crucially, we require it to be sub-linear in the system size $L$.

We also note that states nontrivial under our mixed state definition are also nontrivial under the definition used in Sec.~\ref{sec:general}, since classical disorders form a subset of invertible quantum disorders. However, states that are nontrivial in the sense of Sec.~\ref{sec:general} may not be nontrivial in our current context. 

One consequence of the Def.~\ref{def:channelSPT} is that an SPT in clean system $\rho=|\Psi\rangle\langle\Psi|$, which is nontrivial under any symmetric finite-depth  circuit, may become trivial under a symmetric finite-depth channel. As defined in Def.~\ref{def:SymChannel}, both $\mathcal{H}$ and the ancillary space $\mathcal{A}$ transform faithfully under the average symmetry. For an arbitrary SPT state $|\psi_g\rangle$ protected solely by the average symmetry $G$, one can find a $G$-SPT $|\psi_g^{-1}\rangle^\mathcal{A}$ defined in $\mathcal{A}$, such that the state $|\psi_g\rangle \otimes |\psi_g^{-1}\rangle^\mathcal{A}$ can be prepared from a trivial product state by a finite-depth local unitary with gates that commute with $S(g)=U(g)\otimes U(g)^\mathcal{A}$. This statement is known as the invertibility of SPT states \cite{Kong2014arXiv,Freed2014}. On the other hand, starting from a $G$-SPT $| \psi_g \rangle$, one can always construct a symmetric finite-depth local unitary, which brings $| \psi_g \rangle \langle \psi_g | \otimes | a \rangle \langle a |$ to $\rho_{cl}\otimes | \psi_g \rangle^\mathcal{A} \langle \psi_g |^\mathcal{A}$. After tracing out $\mathcal{A}$, this implies $| \psi_g \rangle$ becomes trivial in the mixed state setting, according to the definition Eq.~(\ref{eq:ASPTdefinition}). This logic also applies to any nontrivial invertible phase (such as the chiral $E_8$ state in $(2+1)d$), as we can also bring the ancillary degrees of freedom into the appropriate inverse state. In this sense, ``gravitational response'' becomes a trivial concept in the mixed state setting.

\subsection{A simple example}
\label{subsec:numerics}

We now discuss an example of nontrivial average SPT phases under the definitions used in this Section. The simplest example is in fact the one discussed in Sec.~\ref{sec:boringboundary}, where one of the $\mathbb{Z}_2$ symmetries in the $\mathbb{Z}_2\times \mathbb{Z}_2$ cluster chain becomes an average symmetry due to a random field perturbation.

One way to characterize the clean cluster model is the nonlocal string order parameter in the ground-state \cite{Pollmann2012s,Pollmann2010}:
\begin{equation}
    \lim_{|n-m|\to\infty} \langle Z_{2m-1} \prod_{k=m}^n X_{2k} Z_{2n+1} \rangle \ne 0.
    \label{eq:stringorder}
\end{equation}
The string order is made out of the symmetry operator $G$ in the middle (but acting in a finite region), multiplied by two local endpoint operators\footnote{ The easiest way of seeing Eq.~(\ref{eq:stringorder}) is by noting that it is equal to $\prod_{k=m}^n Z_{2k-1} X_{2k} Z_{2k+1}$, with $Z_{2k-1} X_{2k} Z_{2k+1}=1$ in the ground-state. Away from the
exactly solvable point, the long range order is no longer perfect, but
the expectation value of the string order remains nonzero -- it is a general feature of 1D SPT phases \cite{Pollmann2012}.}. One can construct a similar string order for the symmetry $K$, i.e. $Z_{2m}\prod_{k=m}^{n-1} X_{2k+1} Z_{2n}$, which also has long range order in the ground-state. For later convenience, we denote a string order associated with a symmetry $K$ by $\mathcal{S}_K$, which is constructed by the symmetry operator $s^K$ (acting in a finite region) in the middle, multiplied by some local endpoint operators: $\mathcal{S}_K=O_K^l s^K O_K^r$.

The topological nature of the cluster SPT is encoded in the symmetry charge of the endpoint operators: in order for the string order associated with symmetry $K$ ($G$) to have long ranged order, its endpoint operators must be odd under symmetry $G$ ($K$). In contrast, in a trivial SPT, e.g. a paramagnetic chain $H_\mathrm{triv}=- \sum_n X_n$, the endpoint operators of a string order with a nonzero ground-state expectation cannot carry any non-trivial charges. These distinct quantized charges indicate the two models must be separated by a phase transition.

We now add the random field
\begin{equation}
    H_{\mathrm{dis}}=-\sum_n h_{2n} Z_{2n},
\end{equation}
where $h_{2n}$'s are onsite potentials distributed uniformly in $[-\delta,\delta]$. The ensemble of ground states now have exact symmetry $K$ generated by Ising spins on the odd-sites, while the Ising symmetry on the even-sites $G$ is only an average symmetry. 

\begin{figure}

\centering
\includegraphics[width=0.5\textwidth]{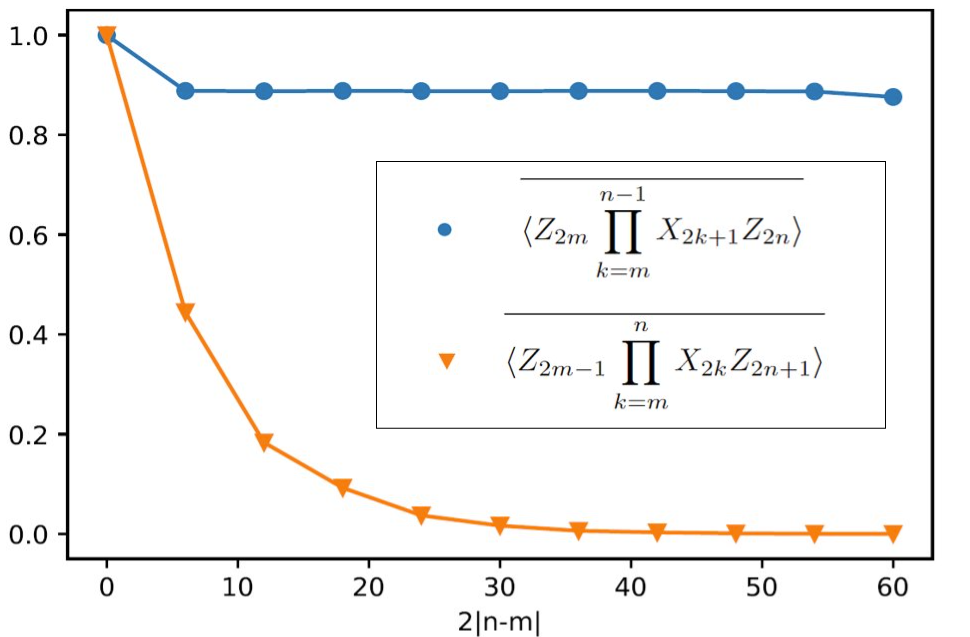}
\caption{The string order parameters associated with $K$ and $G$ respectively, in the presence of disorder with $\delta=0.4$. The overline denotes the ensemble average over 50 samples. The underlying clean model is in the cluster phase, perturbed away from the exactly solvable point. The numerical study is performed using the density matrix renormalization group (DMRG) technique\cite{white1992density,2011Schollw}.}
\label{fig:stringorder}
\end{figure}

One can study the behaviours of the string orders in the presence of this disorder. Since the symmetry $G$ is broken locally by randomness in each realization of disorder, one expects the ensemble average of the string order associated with $G$ to decay exponentially as a function of the length of the string. On the other hand, if the disorder does not close the bulk energy gap (which can be checked given the specific Hamiltonians in Eq.~(\ref{eq:cluster}) and Eq.~(\ref{eq:disorderHamiltonian}), as long as the disorder strength $\delta$ is small compared with the bulk gap), by continuity, we expect that the string order of the unbroken $K$ with non-trivial endpoint operators remains long range ordered. These expectations are confirmed numerically, see Fig.~\ref{fig:stringorder}. One can also add the disorder in Eq.~(\ref{eq:disorderHamiltonian}) to a trivial SPT, e.g. a trivial paramagnet. In contrast, we find numerically that both string orders of $K$ and $G$, with endpoint operators odd under the other symmetry, have no nonzero ensemble average.    

Analogous to the clean case, one may wonder if such a nonzero string order parameter can serve as a characteristic fingerprint of a “non-trivial phase". The answer is yes, as we will show below.
Specifically, we will show that if the ensemble average of the non-trivial string order parameter $\mathcal{S}_k$ (associated with an element $k\in K$) remains long-range ordered, the mixed state $\rho$ cannot be a trivial ASPT. 

\begin{theorem}
\label{thm:string}
Let $\rho$ be a symmetric SRE ensemble in which the non-trivial string order $\mathcal{S}_k$ has long-range order. The trace norm in Eq.~(\ref{eq:ASPTdefinition}) remains non-zero for any choice of symmetric finite-depth local channel.
\end{theorem} 

\emph{Proof}: Consider a symmetric local channel $\mathcal{E}$, constructed as that in Eq.~(\ref{eq:symmetricchannel}). The depth of the circuit $W$ multiplied by the maximum range of each unitary in the circuit is
bounded by some range $R$, which is sub-linear in $L$. Suppose we have a string order parameter of the exact symmetry $\mathcal{S}_k$, with two endpoint operators $O_k^l(x)$ and $O_k^r(y)$ acting in the Hilbert space $\mathcal{H}$ with non-trivial charge under $U(g)$. The length of the string $|x-y|$ is taken to be much larger than $R$. Under the action of the unitary circuit, $\mathcal{S}_k$ is mapped to another string operator $W^\dagger \mathcal{S}_k W$. In the region well separated from the endpoints (with a distance larger than $R$), the string $\mathcal{S}_k$ remains unchanged, as the circuit $W$ commutes with the exact symmetry $S(k)=U(k)\otimes \mathbf{1}^\mathcal{A}$. The endpoint $O_k^l(x)$ is mapped by the circuit to a “local" operator $\Tilde{O}^l_k(x)= W^\dagger O^l_k(x) W$, supported on a region within distance $R$ of $x$. (The discussion for the right endpoint $O_k^r$ is the same, hence omitted hereafter.) Therefore, $W^\dagger \mathcal{S}_k W$ is again a string order parameter associated with the group element $k$.  

An important observation is that the new endpoint operator $\Tilde{O}^{l}_k$ has the same charge under the average symmetry $G$ as $O_k^{l}$, since the circuit $W$ is symmetric: 
\begin{equation}
    \begin{split}
        & S(g)^\dagger \Tilde{O}^l_k S(g) \\
        =& W^\dagger (U(g)\otimes U(g)^\mathcal{A})^\dagger O_k^l (U(g)\otimes U(g)^\mathcal{A}) W \\
        =& W^\dagger U(g)^\dagger O_k^l U(g) W.
    \end{split}
\end{equation}
Remember that $O_k^l$ acts only on $\mathcal{H}$, thus commutes with $U(g)^\mathcal{A}$. Therefore, when we compute the expectation value of $\mathcal{S}_k$ in a trivial ensemble, we have
\begin{equation}
    \mathrm{tr}_\mathcal{H}[\mathcal{S}_k\mathrm{tr}_\mathcal{A}(W \tilde{\rho}_{cl} W^\dagger)]\sim \langle \Tilde{O}_k^l(x) \rangle \langle \Tilde{O}_k^r(y) \rangle=0,
    \label{eq:vanishingstr}
\end{equation}
where $\tilde{\rho}_{cl}$ is a symmetric product state in the enlarge Hilbert space $\mathcal{H}''$, i.e. $\tilde{\rho}_{cl}=\rho_{cl}\otimes | a \rangle \langle a |$, and where $\langle...\rangle$ denotes the expectation value with respect to this state. To get Eq.~(\ref{eq:vanishingstr}), notice that $\mathcal{S}_k=O_k^l s^k O_k^r$ and the string $s^k$ between the endpoints acts trivially on $\tilde{\rho}_{cl}$. We also used the cluster decomposition theorem for two well-separated endpoints. The non-trivial $S(g)$ charge of $\Tilde{O}_k^{l/r}$ then forces the above expectation value to be zero. As a result, if the non-trivial string $\mathcal{S}_k$ is long range ordered in $\rho$, we have 
\begin{equation}
\begin{split}
    &||\rho - \mathcal{E}(\rho_{cl}) ||_1 \\
    &\geq  |\mathrm{tr}_{\mathcal{H}} \mathcal{S}_k [\rho - \mathrm{tr}_\mathcal{A}(W \tilde{\rho}_{cl} W^\dagger)]|/||\mathcal{S}_k||\sim O(1).
\end{split}
\end{equation}
This completes the proof of Theorem~\ref{thm:string}.

Theorem~\ref{thm:string} indicates that an SPT whose protection involves the exact symmetry can not be prepared from a trivial product state. This observation will be made precise below.

\subsection{Domain walls in an ASPT}

\label{subsec:ASPTdomainwall}

{We now show that for quantum disorders, the decorated domain wall picture again emerges naturally within the density matrix description. For simplicity, we use the cluster chain studied above as an example. In this subsection hereafter, we take $g=k=\mathbb{Z}_2$. 
}

For symmetric SRE states, applying the symmetry in a finite but large region (much larger than the correlation length) is equivalent to applying a unitary operator just near the boundary of that region. In $(1+1)d$, the open string $s^k$ effectively only acts near the ends, 
\begin{equation}
    s^k \rho (s^k)^\dagger = \mathrm{tr}_\mathcal{D} U_k^l U_k^r |\psi\rangle \langle\psi| (U_k^l)^\dagger (U_k^r)^\dagger,
\end{equation}
where $s^k$ is fractionalized on the symmetric SRE state $| \psi \rangle$, and $U_k^l$ ($U_k^r$) acts non-trivially only near the left (right) edge. Notice that though the string $s^k$ acts as an identity on the ancillary space $\mathcal{D}$, the operator $U_k^{l/r}$ might acts non-trivially on $\mathcal{D}$. The long range order of $\mathcal{S}_k=O_k^l s^k O_k^r$ implies the expectation value
\begin{equation}
    \langle \psi|O_k^l U_k^l O_k^l U_k^r |\psi\rangle \ne 0,
\end{equation}
for large separations of the two ends. By cluster decomposition theorem, one has
\begin{equation}
    \langle \psi|O_k^l U_k^l |\psi\rangle \ne 0,
    \label{eq:edgecharge}
\end{equation}
and similarly for the right endpoint. As $|\psi\rangle$ is symmetric, when $O_k^l$ is charged under $\tilde{S}(g)$ (like in the case of the cluster chain), the non-vanishing expectation value requires the operator $U_k^l$ also carries a non-trivial $\tilde{S}(g)$ charge. 

Next, instead of the string of the exact symmetry ($s^k$), let us conjugate the density operator $\rho$ by $s^g$ (acts on the Hilbert space $\mathcal{H}$ only), a finite but long string of an average symmetry. Again due to the SRE nature of the purifying state, we have
\begin{equation}
\begin{split}
    s^g\rho (s^g)^\dagger &= \mathrm{tr}_\mathcal{D} s^g |\psi\rangle\langle\psi| (s^g)^\dagger \\
    &= \mathrm{tr}_\mathcal{D} s^g\otimes s^{g\mathcal{D}} |\psi\rangle\langle\psi|(s^g)^\dagger\otimes (s^{g\mathcal{D}})^\dagger \\
    &= \mathrm{tr}_\mathcal{D} U_g^l U_g^r |\psi\rangle\langle\psi|(U_g^l)^\dagger (U_g^r)^\dagger,
    \label{eq:Gfractionalization}
\end{split}
\end{equation}
in which we have to include a corresponding string $(s^g)^D$ acting on $\mathcal{D}$, due to the non-trivial $G$ transformation of the ancillary space (see Eq.~(\ref{eq:averagesym})). A nontrivial result of the cohomology group $H^2(\mathbb{Z}_2\times \mathbb{Z}_2,U(1))$ \cite{Shiozaki2017} states that the $k$ charge of the operator $U_g^{l/r}$ should be identical to the $g$ charge of $U_k^{l/r}$, and is therefore nontrivial. Since the string $s^g$ creates a $g$ domain wall at each endpoint, we thus see that a domain wall of the average symmetry is decorated by a non-trivial charge (i.e. a 0D SPT) of the exact symmetry. This conclusion is a property of the symmetric SRE mixed ensemble $\rho$, which is independent of the specific choice of the purification $| \psi\rangle$.

The above discussion can be generalized to higher dimensions. For example, in $(2+1)d$, instead of string operators, we can consider membrane operators. The details, however, will be more involved and we do not attempt to provide a full exploration. Instead, we shall make the plausible conjecture that, similar to the $(1+1)d$ examples, the group-cohomology result Eq.~\eqref{eq:DDWcoho} for decorated average domain walls captures the classification of bosonic mixed-state SPT phases (with invertible quantum disorders).

We close this section by pointing out a connection between our discussion and Ref.~\cite{deGroot2021arXiv}, which studied mixed state SPT in the context of open quantum systems. The definition of exact and average symmetries in this work mimics the definition of the strong and weak symmetry conditions for quantum channels in Ref.~\cite{deGroot2021arXiv}. The two types of channels (or Lindbladians) there can thus be understood as adiabatically turning on disorder that exactly or averagely preserves the protecting symmetry of an SPT. It was observed in Ref.~\cite{deGroot2021arXiv} that a weakly symmetric channel is insufficient to preserve SPT phases. This, in our language, is the statement that an SPT protected by average symmetry alone is trivialized by disorder, presented in Sec.~\ref{subsec:ASPT}.     

\section{Discussions}

\label{sec:discussion}

We end with some open directions, several of which were also mentioned in previous Sections.

\begin{enumerate}
    \item We have focused on disordered ensembles in which any two states (with different disorder realizations) are adiabatically connected to each other (Def~\ref{def:SREensemble}). This assumption allows us to make controlled arguments, even without assuming weak disorder strength. However, it does leave open the possibility of interesting topological phenomena in disordered ensembles not satisfying this adiabatic assumption. For example, in an Anderson localized insulator, the $U(1)$ charge at each position fluctuates depending on the local chemical potential, so different disorder realizations gives different $U(1)$ charges, and therefore cannot be symmetrically connected to each other. In Sec.~\ref{sec:replicatheory} we also discussed the possibility of sample-to-sample fluctuations that are topological in nature -- such phenomena will certainly require us to go beyond the adiabatic assumption. If such ``topological sample fluctuation'' can indeed happen, it would represent a novel topological phenomenon that intrinsically requires strong disorder.
    
    \item It may also be possible to have ``intrinsically disordered average SPT'' even if the adiabatic assumption in Def.~\ref{def:SREensemble} is kept. As we discussed in Sec.~\ref{sec:generaldecor}, among the set of consistency rules required in the standard decorated domain wall approach, there is one that is not required in the context of average SPT: the domain walls do not need to have consistent Berry phase when moved around, simply because the domain walls are anyway pinned by local disorders and do not move. This leaves open the possibility of average SPT phases not allowed in the clean limit. We will develop the theory of such phases in more detail in a forthcoming work.
    
    \item In Sec.~\ref{sec:boundary} we showed that if the decoration dimension is greater than $(0+1)d$, then the boundary of average SPT state should almost certainly be long-range entangled, with probability approaching $1$ in the thermodynamic limit. It will be desirable, however, to obtain a more direct statement on (averaged) measurable quantities such as correlation functions or inverse energy gap. This is a natural direction for next step.
    
    \item t'Hooft anomaly has been an extremely powerful non-perturbative tool in the study of strongly coupled gapless states of matter, including various conformal field theories that arise in exotic quantum criticality and even compressible states (some recent examples include Refs.~\cite{Ye2021,Else2021}). It is natural to ask whether the disordered version of these states can also be fruitfully studied using the average anomalies.
    
    \item Since we have established the notion of average symmetry-protected topological phase, an immediate question is whether the notion of \textit{average symmetry-enriched topological} (SET) phases can be similarly defined. In particular, are various concepts\cite{Barkeshli2014} in SET well defined for average symmetry? If so, what are their consequences?
    
    \item There are some other scenarios in which mixed states necessarily appear. One is in open quantum systems, where finite depth quantum channels are naturally realized by fast local Lindbladian evolutions \cite{Coser2018,deGroot2021arXiv}.  We therefore expect the results in this work shed light on classification and characterization of SPT phases in open systems. There are several questions remain unclear. For instance, can mixed SPT states arise as steady states of Lindbladian evolutions? Can we formulate a similar field theory, when the Hamiltonian (Lindbladian) is time-dependent? These open questions are left to future study. 
\end{enumerate}

\begin{acknowledgements}

We thank Yushao Chen, Lei Gioia, Meng Guo, Itamar Kimchi, Rahul Nandkishore, Sri Raghu, Shengqi Sang, Cenke Xu and Jianhao Zhang for helpful discussions. RM is especially grateful to his officemate Matthew Yu, a master of algebraic topology, for patiently answering his numerous questions. DMRG simulations were performed using the
TenPy tensor network library\cite{2018DMRG}. RM acknowledges  supports  from  the  Natural Sciences and Engineering Research Council of Canada(NSERC)  through  Discovery  Grants. Research at Perimeter Institute is supported in part by the Government of Canada through the Department of Innovation, Science and Industry Canada and by the Province of Ontario through the Ministry of Colleges and Universities.

\end{acknowledgements}

\bibliography{Ref.bib}

\clearpage
\onecolumngrid
\appendix

\section{Atiyah-Hirzebruch spectral sequence for Class AII}
\label{app:SSS}
In order to define a fermionic theory in symmetry class AII, one should equip the space-time manifold with a $\mathrm{pin}_+^{\tilde{c}}$ structure. In $(d+1)$-dimension, the structure group fits in the short exact sequence:
\begin{equation}
\begin{tikzcd}
1 \arrow[r] & U(1) \arrow[r] & \mathrm{pin}_+^{\tilde{c}} \arrow[r] & O(d+1) \arrow[r] & 1,
\end{tikzcd}
\label{eq:classAII}
\end{equation}
with the reflection element in $O(d+1)$ squares to $1$ in Euclidean signature and acts on $U(1)$ by complex conjugation. For our purpose, we calculate the cobordism $\Omega^\bullet_{\mathrm{pin}_+^{\tilde{c}}}$ using an AHSS with $E_2$ page given by $E_2^{p,q}=H^p(B\mathbb{Z}_2;\Omega^q_{\mathrm{spin}^c}(\mathrm{pt}))$, with the coefficient group twisted appropriately by $\mathbb{Z}_2$. For example, the IQH root state and the $E_8$ root state are both time reversal odd, so $\mathbb{Z}_2$ acts on their corresponding elements in $\Omega^\bullet_{\mathrm{spin}^c}$ non-trivially. On the other hand, $\mathbb{Z}_2$ acts on the $U(1)$ charge trivially. The $E_2$ page in low degree is given by
\begin{equation}
    \begin{array}{c|cccccc}
    2 U(1)^\mathcal{T} & 2 (-1)^{\mathbb{Z}_2} & & 2 (-1)^{\mathbb{Z}_2t^2} & & 2(-1)^{\mathbb{Z}_2 t^4} & \\
     0 & & & & & &\\
    U(1) & U(1) & (-1)^{\mathbb{Z}_2 t} &  & (-1)^{\mathbb{Z}_2 t^3} & & (-1)^{\mathbb{Z}_2 t^5}\\
     0 &  &  &  &  & & \\
    U(1)^\mathcal{T} & (-1)^{\mathbb{Z}_2} &  & (-1)^{\mathbb{Z}_2 t^2} &  & (-1)^{\mathbb{Z}_2 t^4} & \\ \hline
    & 0 & 1 & 2 & 3 & 4 & 5 \, ,
    \end{array}
\end{equation}
in which $U(1)^\mathcal{T}$ indicates the coefficient twisted by time reversal. $t$ is the generator of the cohomology ring $H^\bullet(B\mathbb{Z}_2;\mathbb{Z}_2)=\mathbb{Z}_2[t]$ with $t$ in degree one. In this spectral sequence only the $d_3$ differential can possibly be non-trivial. Given\cite{2019GaiottoJF,2021Yu}
\begin{equation}
d_3=(-1)^{(\mathrm{Sq}^2+t\cdot \mathrm{Sq}^1 +t^2)}\circ \beta,    
\end{equation}
for $\mathcal{T}^2=-1$ when acting on fermions, one can see that the $d_3$ differential vanishes for elements with total degrees up to 4. Here $\beta$ is the Bockstein of the following sequence in cohomology:
\begin{equation}
\begin{tikzcd}
1 \arrow[r] & \mathbb{Z}_2 \arrow[r, "(-1)^x"] & U(1) \arrow[r, "x^2"] & U(1) \arrow[r] & 1
\end{tikzcd}
\end{equation}
such that $\beta: H^n(B\mathbb{Z}_2;U(1))\to H^{n+1}(B\mathbb{Z}_2;\mathbb{Z}_2)$. Physically, the vanishing of differential means the decorations we discussed in Sec.~\ref{sec:fermionicexample} are consistent. The calculation also agrees with the $\mathbb{Z}_2^3$ classification of class AII TIs in three spatial dimensions \cite{WangPotter2014}.

\end{document}